\documentclass[iop,revtex4]{emulateapj}

\usepackage{graphicx}
\usepackage{float}
\usepackage{amsmath}

\usepackage{enumitem}
\setlist{parsep=0pt,listparindent=\parindent}
    {\pagebreak[4]\global\pdfpageattr\expandafter{\the\pdfpageattr/Rotate 90}}%
    {\pagebreak[4]\global\pdfpageattr\expandafter{\the\pdfpageattr/Rotate 0}}%

\newcommand{\DcHR}{%
  \ensuremath{\delta \langle M_B^{\mathrm{corr}}\rangle_{\mathrm{SF}}}}
\def\Rvlowbias{0.059}

\def\Rvlowbiassig{2.6}
\def\Rvlowbiassyssig{2.4}
\def\Rvlowbiaserrfull{0.025}

\def\Rvmidbias{0.029}

\def\Rvmidbiassig{1.2}
\def\Rvmidbiassyssig{1.1}
\def\Rvmiddiscrep{2.6}
\def\Rvmidbiaserrfull{0.027}

\def\Rvhighbias{0.013}

\def\Rvhighdiscrep{3.2}
\def\Rvhighbiaserrfull{0.030}

\def\SALTbias{0.000}

\def\SALTbiassyssig{0.0}
\def\SALTdiscrep{2.3}
\def\SALTbiaserrfull{0.018}

\def\numJLA{179}
\def\numJLAfull{207}
\def\numRiess{157}
\def\numRiessfull{187}
\def\SSFR{log($\Sigma_\textrm{SFR}$)}

\newcommand{\JHU}{Department of Physics and Astronomy, The Johns Hopkins University, Baltimore, MD 21218.}
\newcommand{\STScI}{Space Telescope Science Institute, Baltimore, MD 21218.}
\newcommand{\Kavli}{The Kavli Institute for Cosmological Physics, University of Chicago, Chicago, IL 60637, USA.}

\altaffiltext{1}{\JHU}
\altaffiltext{2}{\STScI}
\altaffiltext{3}{\Kavli}

\begin{document}

\title{Reconsidering the Effects of Local Star Formation
  On Type Ia Supernova Cosmology}
\author{David O. Jones\altaffilmark{1}, Adam G. Riess\altaffilmark{1,2}, Daniel M. Scolnic\altaffilmark{3}}

\begin{abstract} 

Recent studies found a correlation with $\sim$3$\sigma$ significance between the local
star formation measured by GALEX in Type Ia supernova (SN\,Ia) host galaxies and the
distances or dispersions derived from these SNe.  We search for these
effects by using data from recent cosmological analyses to greatly increase the SN\,Ia sample; we include
\numJLA\ GALEX-imaged SN\,Ia hosts with distances from the JLA and Pan-STARRS SN\,Ia cosmology samples and
\numRiess\ GALEX-imaged SN\,Ia hosts with distances from the \citet{Riess11} H$_0$
measurement.  We find little evidence that SNe\,Ia in locally
star-forming environments are fainter after light curve correction than SNe\,Ia in locally
passive environments.  We find a difference of 
\SALTbias$\pm$\SALTbiaserrfull\ (stat+sys) mag for SNe fit with SALT2 and
\Rvmidbias$\pm$\Rvmidbiaserrfull\ (stat+sys) mag for SNe fit with
MLCS2k2 (R$_V = $ 2.5), which suggests that proposed changes to recent
measurements of H$_0$ and $w$ are not significant and numerically smaller than the parameter measurement uncertainties.  We
measure systematic uncertainties of $\sim$0.01-0.02 mag by
performing several plausible variants of our analysis.  We find
the greatly reduced significance of these distance modulus differences compared to \citet{Rigault13}
and \citet{Rigault15} result from two improvements with fairly equal effects, our larger sample
size and the use of JLA and \citet{Riess11} sample
selection criteria.  Without these improvements, we recover the results
of \citet{Rigault15}.  We find that both populations have more
similar dispersion in distance than found by \citet{Rigault13},
\citet{Rigault15}, and \citet{Kelly15}, with slightly smaller
dispersion for locally passive (\SSFR\ $<-2.9$ dex)
SNe\,Ia fit with MLCS, the opposite of the effect seen by \citet{Rigault15}
and \citet{Kelly15}.  We caution that measuring the local environments of SNe\,Ia in the
future may require a higher-resolution instrument than GALEX and that
SN\,Ia sample selection has a significant effect on
local star formation biases.



\end{abstract}

\section{Introduction}

Type Ia supernovae (SNe\,Ia) have been 
a key component in measuring the dark energy equation of state,
$w$, with $\lesssim$6\% uncertainty \citep{Betoule14} and the
Hubble Constant, H$_0$, with 3.3\% uncertainty (\citealp{Riess11};
hereafter R11).  With such small error budgets, unknown 
systematic uncertainties affecting SNe\,Ia shape- and color-corrected
absolute magnitudes
could have serious consequences for our understanding of
dark energy, neutrino properties, and the global geometry of space.

Although SNe\,Ia remain accurate distance indicators with $\sim$10\%
uncertainty per SN, there are
concerns about their ability to remain standardizable in galaxies that
vary in mass, metallicity, star formation, age, and dust properties 
(e.g. \citealp{Sullivan10,Rigault13,Johansson13,Childress13}).  Even
a small dependence of SN\,Ia luminosities on host galaxy properties may have a non-negligible
effect on $w$ due to the redshift evolution of
galaxies or differences in sample selection.  Such an effect could also
bias H$_0$ due to the different demographics of Cepheid host galaxies compared to SN\,Ia
hosts.  The lack of detection of such an effect at $>$3$\sigma$ with samples
of $\sim$10$^2$ SNe suggests that such effects are $\lesssim$
$\frac{10\%}{\sqrt{100}} \times 3 \lesssim 0.06$ mag, or that they result from galaxy
  properties that are difficult to measure robustly.  These
  investigations are hampered by an inability to define the nature of
  the SN\,Ia correction \textit{a priori}, complicating the interpretation of
  the significance of the correlations found \textit{a posteriori}.
  If enough sources for a possible correlation are examined, a
  3$\sigma$ result will always be found.


The first widely accepted effect of host galaxy properties on
SNe\,Ia was confirmed by the detection of a $\sim$0.07 mag difference
in mean corrected magnitude of SNe\,Ia with host masses
$>$10$^{10}M_{\odot}$.  Identified by several independent studies including \citet{Lampeitl10},
\citet{Sullivan10}, and \citet{Kelly10}, this effect has now been
detected at $>$5$\sigma$ by \citet{Betoule14} with a sample of 740
SNe\,Ia.

Because it is unclear how the physics of a SN\,Ia distances could
depend on its host galaxy mass, the most likely explanation is 
that host galaxy mass is merely tracing
another physical property that could affect SN luminosity, such as metallicity, stellar
age, or dust.  \citet{Dominguez01} suggested that progenitor
metallicity could affect the SN luminosity by changing the Carbon-Oxygen ratio in the progenitor
white dwarf, thus resulting in a lower Nickel mass synthesized in the
explosion.  \citet{Hayden13}
found that a correction using a star formation-based metallicity indicator reduced Hubble
diagram residuals more than a simple host mass correction.
\citet{Childress13} found that dust and stellar age are also plausible
explanations because they evolve with host galaxy mass.

Different SN\,Ia progenitor ages could
also exhibit systematic differences in corrected magnitude due to
the effects of metallicity or explosion mechanism on $^{56}$Ni production \citep{Maoz14}. \citet{Childress14} 
suggested that progenitor age could be the source
of the host mass step, as older progenitors preferentially occur in
non star-forming host galaxies.  Because progenitor age evolves with redshift, \citet{Childress14}
modeled a potential redshift-dependent bias in cosmological analyses.



SN\,Ia light curve fitters may also
create biases by assuming a universal relationship between color and absolute magnitude,
independent of the dust composition of different SN\,Ia hosts.
Some preliminary evidence has supported these ideas; \citet{Scolnic14}
found that the correlation between SN\,Ia color and
absolute magnitude has two different slopes for bluer and redder SNe,
which may in part be due to dust properties.

\begin{figure*}
\includegraphics[angle=0,width=7.in]{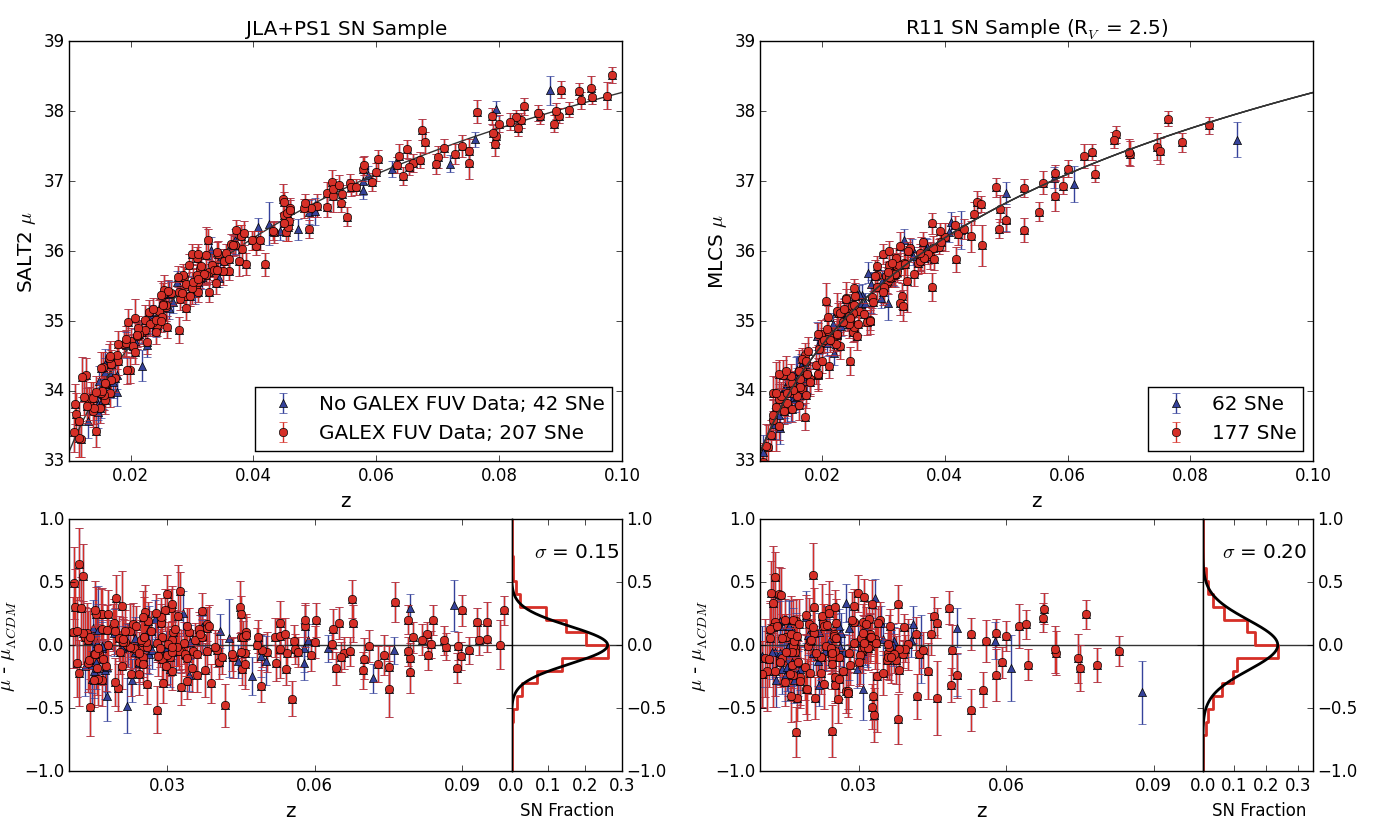}
\caption{Hubble diagrams and Hubble residual diagrams for the JLA+PS1
  sample (SALT2 light curve fitter; left) and the R11
  sample (MLCS light curve fitter with R$_V = 2.5$; right), with GALEX FUV-imaged hosts in red and
  SNe without GALEX FUV host images in blue.  Out of a 
total of 249 SNe in the JLA+PS1 sample, 207 were imaged by GALEX
within 0.55 degrees of field center.  In the R11 sample, 177 out of
239 SNe fit with R$_V = 2.5$ had GALEX FUV images.  The MLCS data have slightly higher scatter,
but both samples have intrinsic dispersions $\lesssim$0.2.}
\label{fig:hubble}
\end{figure*}

If the host mass step is indicative of one or more of these biases, 
galaxy properties in the vicinity of SN explosions could be more
strongly correlated with SN corrected magnitude than properties of the galaxies
as a whole.
Three recent studies used $\sim$60$-$85 nearby SNe\,Ia to look at such properties and found
that they affect the distances derived from 
SNe\,Ia.  \citet{Rigault13} and \citet{Rigault15} found a correlation
between local star formation and SN\,Ia Hubble residuals from the
Nearby Supernova factory \citep{Aldering02} and the CfA SN survey
\citep[hereafter H09]{Hicken09} by using the
local star formation rate density ($\Sigma_{SFR}$) to separate SNe\,Ia into
those with locally passive (SN\,Ia$\epsilon$) and locally star-forming 
(SN\,Ia$\alpha$) environments.  \citet{Rigault15} (hereafter R15) found a mean
difference in Hubble residuals between SNe\,Ia$\epsilon$ and
Ia$\alpha$ (hereafter referred to as the LSF step)
of $\sim$0.09$-$0.17 mag at 2-4$\sigma$ significance with different light curve fitters.  

The fraction of SNe\,Ia$\epsilon$ is
different in the nearby Cepheid-calibrated SN\,Ia sample compared to the
Hubble-flow SN\,Ia sample, and R15 found that SNe\,Ia$\epsilon$ have
mean corrected magnitudes $\sim$0.15 mag brighter than
SNe\,Ia$\alpha$ when fit with the MLCS light curve fitter 
and assuming the same R$_V$ as the R11
H$_0$ baseline analysis.  They derived a correction to H$_0$:

\begin{equation}
 \label{eq:H0}
  \log(H_0^{corr}) = \log(H_0)
  - \underbrace{
    \frac{1}{5}(\psi^{HF}-\psi^{C})\times \DcHR,}_{\text{LSF~bias correction}}
\end{equation}

\noindent where $\psi^{HF}$ is the fraction of SNe\,Ia$\epsilon$ in the
Hubble-flow SN sample and $\psi^{C}$ is the fraction of SNe\,Ia$\epsilon$ in
the Cepheid-calibrated sample.  $\DcHR$ is the LSF step of 0.155 mag.
By estimating $\psi^{HF}$ (52.1$\pm$2.3\%) and $\psi^{C}$ (7.0\%), R15 
estimate that the true value of H$_0$ is reduced by $\sim$3\%.

R15 also found that SNe
in highly star-forming regions fit by MLCS \citep{Jha07,Riess96} have lower dispersion in
their Hubble residuals than
SNe in locally passive environments.  \citet{Kelly15} came to the same conclusion by
examining SNe\,Ia with high local star formation
(Their $\Sigma_{\textrm{SFR}}$ boundary is $\sim$0.7 dex
higher than the R15 Ia$\epsilon$/Ia$\alpha$ cut-off).  R13 first found this
effect using the SALT2 light curve fitter \citep{Guy07}, but they could not reproduce
this result with H09 data.

\begin{deluxetable*}{p{0.65in}ccccccccc}
\tablewidth{0pt}
\tablewidth{0pt}
\tablecolumns{10} 
\tabletypesize{\scriptsize}
\tablecaption{Studies using local SF data}
\renewcommand{\arraystretch}{1}
\tablehead{&&\multicolumn{3}{c}{SALT2}&&\multicolumn{4}{c}{MLCS}\\
\cline{3-5} \cline{7-9}\\
&SN Surveys&SNe&$\mu_{\textrm{version}}$&$\beta$&&SNe&$\mu_{\textrm{version}}$&P($A_V$)&R$_V$\\}
\startdata
Rigault+13&SNfactory\tablenotemark{a}&82&G07\tablenotemark{b}&\nodata\tablenotemark{c}&&\nodata&\nodata&\nodata&\nodata\\
Rigault+15&CfA3&77&G07\tablenotemark{b}&2.48$^{+0.10}_{-0.12}$&&84&v0.06&$e^{-A_V/0.457}$&1.7,2.5,3.1\\
Kelly+15&LOSS\tablenotemark{d},CfA2-4,CSP&\nodata&\nodata&\nodata&&61&v0.07\tablenotemark{e}&$e^{-A_V/0.3}\ast\mathcal{N}$($\sigma=0.02$)\tablenotemark{f}&1.8,3.1\\
This Work&CfA1-4,CSP,CT\tablenotemark{g},SDSS,SNLS,PS1&187&G10\tablenotemark{h}&$3.097\pm0.062$&&154&v0.06&$e^{-A_V/0.457}$&2.0,2.5,3.1\\
\enddata
\label{table:samples}
\tablenotetext{a}{\citet{Aldering02}.}
\tablenotetext{b}{\citet{Guy07}.}
\tablenotetext{c}{The value of $\beta$ was blinded in \citet{Rigault13}.}
\tablenotetext{d}{The Lick Observatory Supernova Search \citep{Li11}.}
\tablenotetext{e}{MLCS v0.07 used new spectral templates
 from \citet{Hsiao07}.  This version was implemented in the SuperNova
  ANAlysis software \citep[SNANA]{Kessler09b}.}
\tablenotetext{f}{An exponential convolved with a normal distribution
  having $\sigma=0.02$ mag.}
\tablenotetext{g}{Calan/Tololo \citep{Hamuy96}.}
\tablenotetext{h}{\citet{Guy10} had improved uncertainty propagation
  and handling of residual scatter, a new SN\,Ia
  spectral energy distribution regularization scheme, and used a larger training sample with
  higher-$z$ SNe (see their Appendix A for details).}
\end{deluxetable*}

Both R15 and \citet{Kelly15} used GALEX FUV data to measure the star
formation rate within a few kpc of SNe\,Ia positions.  In this work, we use a similar
method to examine whether the significance of the LSF step and reduced dispersion from SNe in locally star-forming
host galaxies is reduced when we use the most current vintage SNe\,Ia distance
estimates, use a much larger sample size, and vary the priors and assumptions used in the original analyses.

Table \ref{table:samples} shows the sizes of the SN samples used in
\citet{Rigault13}, R15, \citet{Kelly15}, and this work, along with the light curve
fitters used, the SALT2 color parameters, and the MLCS prior on
$A_V$.  \citet{Rigault13} used 82 SNfactory SNe with star formation
estimated using local H$\alpha$ from integral field spectroscopy.
\citet{Rigault15} used $\sim$100 SNe from the CfA3 sample of H09, with
$\sim$80 passing GALEX sample cuts.  \citet{Kelly15} used several
surveys but made strict sample cuts and only used SNe with Hubble residuals $<$ 0.3 mag, which
would amount to a $\sim$1.3$\sigma$ cut for R11 data.


By using a sample size $\sim$2-3 times as large as those in the
analyses above, we hope to obtain a robust measurement of the magnitude and uncertainty
of the effect of local star-formation on SN\,Ia corrected magnitudes.
\S2 presents our sample selection,
and \S3 discusses our LSF step and dispersion analysis.  In \S4 and \S5 we present our results and discuss their significance,
and our conclusions are in \S6.

\section{Data}

We used two samples of SNe for this analysis, one from the R11
measurement of H$_0$ and the other from the dark
energy equation of state measurements of \citet{Betoule14} and
Pan-STARRS (PS1; \citealp{Rest14,Scolnic14b}; Scolnic et al. 2015, in prep).  These two
samples rely on many of the same SNe, but R11 use the MLCS light curve
fitter to perform their baseline analysis while \citet{Betoule14}
and PS1 use SALT2 \citep[version 2.4]{Guy10,Betoule14}.  Each sample
is $\sim$2-3 times as large as the R15 and \citet{Kelly15} GALEX-imaged host samples and removes the
possibility of biases between our sample and the samples used in the
most recent measurements of cosmological parameters.

\subsection{\citet{Riess11} SNe}

The H$_0$ determination of R11 use the MLCS2k2 light curve
fitter for their baseline analysis.  We use only their MLCS2k2
distance moduli, as JLA+PS1 consists of a larger SALT2-fit SN\,Ia sample with
more robust light curve cuts and an updated SALT2 model and color parameter, $\beta$.
The R11 sample consists of 140 SNe between
0.023 $<z<$ 0.1 from \citet{Hicken09b} and \citet{Ganeshalingam10}.
As one of the variants in their systematics section, R11 extend the lower
bound of the redshift range to 0.01 after making peculiar velocity
corrections (using results from \citealp{Neill07}
and the \citealp{Pike05} dipole), giving 240 SNe (with peculiar velocity uncertainties
added in quadrature to the distances).  Adopting this redshift range raises H$_0$ by 0.8
km s$^{-1}$ Mpc$^{-1}$, or 0.26$\sigma$.  We adopt this lower redshift
limit of 0.01 as it
allows us to add more SNe\,Ia to our sample, although these nearby SNe
have less weight in the
likelihood approach outlined in \S3 due to their included peculiar velocity uncertainties.  In \S\ref{sec:lsfbias}, we examine the effect of
restricting the redshift range to $z>0.023$.  R11 remove 4$\sigma$
Hubble diagram outliers but make no sample cuts based on light curve
shape, $A_V$, or MLCS $\chi^2$.

MLCS2k2 determines the distance modulus for each SN\,Ia by fitting for
the light curve shape and extinction assuming an extinction prior and
a value for the total-to-selective extinction ratio, $R_V$.  Common
extinction priors include exponential distributions ($e^{-A_V/\tau}$;
see Table \ref{table:samples}), exponential
distributions convolved with gaussians, a flat prior (with or without
negative $A_V$ allowed), and priors based on
host galaxy information.  R11 consider the latter two priors in their
systematic uncertainty analysis, and use an exponential with scale
length 0.457 mag for their baseline analysis.
R11 consider dust reddening laws with 
$R_V = $ 1.5, 2.0, 2.5, and 3.1, using $R_V
= 2.5$ for their baseline analysis.  $R_V = 3.1$ corresponds to
the Milky Way reddening law \citep{Cardelli89}.  We exclude $R_V =
1.5$ from our analysis as such a low value is not typically used in
cosmological analyses (e.g. \citealp{Kessler09} adopt R$_V =
2.18\pm0.5$ for SDSS cosmology); although highly reddened SNe\,Ia tend to favor
low values of R$_V$ \citep{Burns14}, these SNe are usually excluded
from samples used to measure cosmological
parameters.  H09, for example, use only SNe with $A_V < 0.5$.

We queried GALEX\footnote{\url{http://galex.stsci.edu/GalexView/}} for FUV images at the locations of these SNe, keeping
only those with a angular distance from the field of view center (FOV radius) $<$ 0.55 deg to ensure accurate
photometry and avoid reflection artifacts and distortion of the PSF near
the detector edge.  Of the 240 SNe used in R11, we found \numRiessfull\
SN host images meeting this criterion, \numRiess\ of which remained
after the sample cuts described in \S3.  A Hubble diagram of the R11
SN\,Ia sample is shown in Figure \ref{fig:hubble}.  There is less than
0.01 mag difference in mean Hubble residual between the full sample
and the GALEX-detected sample.  No bias is expected for SNe with
GALEX host images.

\subsection{\citet{Betoule14} and Pan-STARRS SNe}

The most recent measurements of $w$
\citep{Betoule14,Rest14} use the SALT2 light curve fitter, 
and compute distance moduli using the equation \citep{Tripp98}:

\begin{equation}
\mu = m_B^{\ast} + \alpha \times X_1 - \beta \times C - M,
\end{equation}

\noindent where $\mu$ is the SN
distance modulus, $m_B^{\ast}$ 
is the peak SN $B$ band magnitude, $X_1$ is the light curve stretch parameter,
and $C$ is the light curve color parameter.  SALT2 adopts a linear
relation between SN\,Ia color and luminosity with no prior.  For
consistency with the JLA cosmological analysis, we only use the SALT2
fitter with these data.


The nuisance parameters $\alpha$, $\beta$, and
$M$ (in this analysis, a single value independent of host galaxy mass)
are simultaneously fit to the full supernova sample.  
In recent work, the value of $\beta$ has risen due to changes in the SALT2 model and larger SN\,Ia
samples.  The value found by \citet{Betoule14} is $\beta = 3.102\pm0.075$,
a difference of $\sim$0.6 relative to the H09 value of
2.48$^{+0.10}_{-0.12}$ (used by R15).  This could have an important impact on
measuring the LSF step, which we discuss further in \S5.1.  In this analysis, we simultaneously fit JLA and
PS1 data together, finding $\beta = 3.097\pm0.062$.  In contrast to
\citealp{Betoule14} and following the R15 claim that the LSF step
replaces the host mass step, we did not apply the host mass step in deriving
this value.



We limited the \citet{Betoule14} Joint Light-curve Analysis
(JLA) to $z<$ 0.1 because the large GALEX PSF
makes the star formation measurement non-local with FWHM $\sim$8 kpc.  This
low-$z$ sample includes data from low-redshift
surveys such as CfA1-3 \citep{Riess99,Jha06,Hicken09b}, the
Carnegie Supernova Project \citep{Hamuy06,Stritzinger11} and Calan/Tololo \citep{Hamuy96}, and surveys
extending to higher $z$ such as SDSS \citep[25 SNe after sample cuts]{Kessler09} and SNLS
\citep[no SNe after sample cuts]{Conley11}.  We added low-$z$ CfA4 SNe from \citet[used in the
PS1 analysis]{Hicken12}, PS1 SNe from \citet{Rest14} and
the upcoming 4-year PS1 cosmological analysis (12 SNe after sample cuts; Scolnic et al. 2015, in
prep).  For both JLA and PS1, peculiar velocities are corrected
following \citet{Neill07} based
on the \citet{Hudson04} model.

The cuts applied to these data are
listed in \citet[their Table 6 and Appendix A]{Betoule14}.  They make
light curve shape, color, and SALT2 fit probability cuts (requiring a fit
probability $>$0.01).  We applied these same cuts to
PS1 SNe, and removed 3.5$\sigma$ outliers from the full sample,
including the 4 $>$3$\sigma$ outliers removed by \citet{Betoule14}.

The JLA and PS1 samples with 0.01 $<z<$ 0.1 contain
a total of 249 SNe.  \numJLAfull\ were found in GALEX with FOV radius
$<$0.55 deg and \numJLA\ remained after the sample cuts
described in \S3.  We found no significant difference ($<$0.01 mag) between mean
Hubble residual of the GALEX-detected sample and the full sample.

Figure \ref{fig:hubble} shows a Hubble diagram for SNe in both samples
with and without GALEX imaging.  Our cosmological fits used
$\Omega_M = 0.3$, $\Omega_{\Lambda} = 0.7$, $w = -1$, H$_0 = 70$ km
s$^{-1}$ Mpc$^{-1}$ and determined the absolute SN
magnitude $M$ from a least squares fit to the Hubble residuals.

\begin{figure*}
\includegraphics[angle=0,width=7.in]{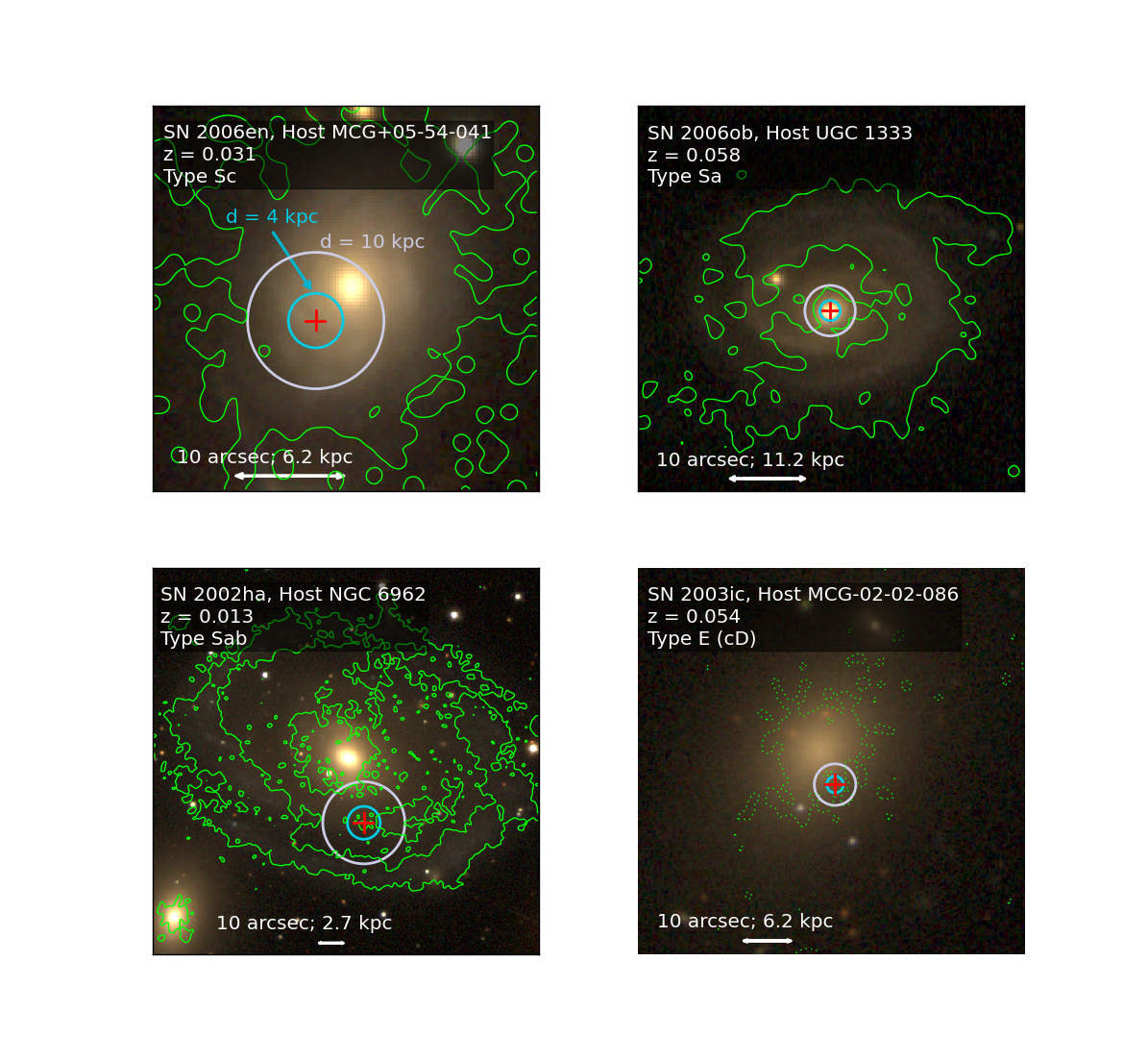}
\caption{Four host galaxies from our sample in SDSS $gri$
  images, with smoothed GALEX FUV contours marking the star-forming regions
  (\SSFR\ $>-2.9$) and the SN\,Ia positions marked in red.  Two apertures are
  overlaid, the local aperture size from R15 (4 kpc diameter)
and the local aperture size from \citet[10 kpc diameter]{Kelly15}.  We
assumed $A_{FUV} = 2.0$ for the three star-forming galaxies.  For the
passive host of SN 2003ic, none of the galaxy would be considered
locally star-forming for $A_{FUV} = 0$, but we show dotted contours to
indicate the effect of assuming 2 mags extinction.  The 4 kpc diameter aperture
appears to be a good approximation for the local star-forming
environment while the 10 kpc aperture extends well beyond the
local star formation environment for SN 2002ha and encompasses most of
the galaxy for SN 2006en.  Both the size of
our local apertures and our prior on $A_{FUV}$
have an important effect on our results, so we vary both
in our systematic error analysis.}
\label{fig:galexlocal}
\end{figure*}


\section{Measuring the Star Formation Density}
\label{sec:R15comp}

\begin{deluxetable*}{lcccc}
\tablewidth{0pt}
\tablewidth{0pt}
\tablecolumns{9} 
\tabletypesize{\scriptsize}
\tablecaption{SN Selection Cuts}
\renewcommand{\arraystretch}{1}
\tablehead{&
\multicolumn{1}{c}{JLA+PS1} &
\multicolumn{3}{c}{R11}\\
No. SNe\,Ia&&R$_V=$2.0&R$_V=$2.5&R$_V=3.1$}
\startdata
Initial Sample&249&240&239&237\\
GALEX FUV data exist&212&189&188&187\\
FOV radius $<0.55$ deg&207&181&180&179\\ 
Global SFR known&207&178&177&176\\
Inclined SNe Removed&179&157&156&155\\

\enddata
\label{table:samplecuts}
\end{deluxetable*}

R15 used the following procedure to measure the \textit{local} star formation
density, $\Sigma_{\textrm{SFR}}$, and its relation to SN distance estimates.  We summarize the principal steps below and describe the differences in
our analysis in \S\ref{sec:procedure}.  \S\ref{sec:analysissyserr} discusses our systematic error
treatment.  Table \ref{table:samplecuts} gives a summary of the
quality cuts applied to our SN\,Ia sample and the number of SNe remaining after
each cut.

\begin{enumerate}
\item R15 measured GALEX FUV aperture
photometry at the location of the SN using a 4 kpc aperture diameter.
They applied Milky Way dust corrections from \citet{Schlegel98}, where the FUV
extinction $A_{FUV}$ is $7.9 \times E(B-V)$ (R15; \citealp{Cardelli89}).
\item The photometry was corrected for host galaxy extinction in the
FUV based on the measured FUV$-$NUV colors, which were converted to
extinction using the relation from \citet{Salim07}.
A Bayesian prior of $A_{FUV} = 2.0\pm0.6$ for star-forming
galaxies was also applied (the final $A_{FUV}$ was a weighted 
mean of the prior and the measured A$_{FUV}$).  R15 made no dust correction for
passive galaxies.

To determine whether each galaxy was globally star-forming or passive,
they used $\Sigma_{\textrm{SFR}}$ measurements from
\citet[$\Sigma_{\textrm{SFR}} > -10.5$ is star-forming]{Neill09}, who
fit synthetic templates to the SN host UV+optical spectral energy 
distributions (SEDs).  Because \citet{Neill09} SED
fits were unavailable for $\sim$40\% of their hosts, R15 
used morphology for these, treating galaxy types Sa and later as star-forming
(a less accurate method).
\item To minimize the effects of locally passive regions projected on
  top of locally star-forming regions (see R15, Appendix B.2), R15 removed SNe with
host inclination angles $>$80$^{\circ}$ from their sample.
\item Based on their photometric and dust correction uncertainties, 
R15 calculated the probability of a SN\,Ia being
above (P(Ia$\alpha$)) or below (P(Ia$\epsilon$)) the
log($\Sigma_{\textrm{SFR}}$) $=-2.9$.
\item R15 used a maximum likelihood approach (outlined in
  \S\ref{sec:maxlike}) to
determine the difference in corrected magnitude and dispersion between
SNe\,Ia$\alpha$ and Ia$\epsilon$.
\end{enumerate}

\subsection{Our Analysis}
\label{sec:procedure}

We largely used the same methodology as R15, but improved the following
aspects of the analysis:
\begin{enumerate}
\item We used the \citet{Schlafly11} dust corrections instead of the
  \citet{Schlegel98} corrections used by R15, resulting in a $\sim$14\%
  reduction in our extinction values.
\item We used SDSS NUV$-$$r$ color instead of morphology as a diagnostic of global SFR when
  UV+optical SED fits were unavailable.
\item For SNe outside the isophotal radii of their host, we did not
  make a dust correction as we expect these SNe to be minimally
  affected by extinction.
\item We made a slightly more conservative inclination cut, removing
  galaxies with inclinations $>$70$^{\circ}$.
\item Using our maximum likelihood model, we fit for both SN\,Ia$\alpha$ and SN\,Ia$\epsilon$ dispersion
  when determining the LSF step to allow for the possibility that
  these two quantities are 
  significantly different and affect the magnitude of the step.
\end{enumerate}
We discuss our changes and methodology in further detail below.  However, these
changes have only minor significance on our results (see \S\ref{sec:checks}).  Our method of
maximum likelihood estimation for calculating the LSF step is described in detail in the Appendix.

\subsubsection{FUV Aperture Photometry}

We used the same baseline 4 kpc
aperture diameter as R15 for our photometry but corrected for Milky Way FUV extinction
using the \citet{Schlafly11} dust
corrections\footnote{\url{http://irsa.ipac.caltech.edu/applications/DUST/}}
instead of the \citet{Schlegel98} corrections used by R15.  \citet{Schlafly11} derive a
$\sim$14\% correction for the \citet{Schlegel98} dust maps based on the
expected vs. measured colors of SDSS stars.  

Using GALEX to estimate local star formation, as in \citet{Rigault15}
and \citet{Kelly15} is complicated by the large GALEX PSF, 5.4$''$
full width at half maximum (FWHM) in
the NUV and 4.5$''$ in the FUV, which serves as a lower limit to the size of
the local region that we can measure.  \citet{Kelly15} used a 10 kpc
aperture diameter to measure local star formation, while
\citet{Rigault15} used a 4 kpc diameter.  We adopt the R15 4 kpc
diameter in this work.


Figure \ref{fig:galexlocal}
shows representative hosts from our sample with FUV-based
\SSFR\ $\geq -2.9$ contours to demonstrate the
size of these apertures relative to their
star-forming regions.  A 4 kpc
aperture appears to be a reasonable approximation to the local SN\,Ia
environment in these cases, while a 10 kpc aperture radius encompasses 
the majority of the SN 2006en host.  In the case of SN 2002ha, it is
unclear whether either aperture is small enough to capture the star
formation environment at the SN location.






\subsubsection{Host Galaxy Extinction Correction}
\label{sec:host ext}

\begin{figure}
\includegraphics[angle=0,width=3.4in]{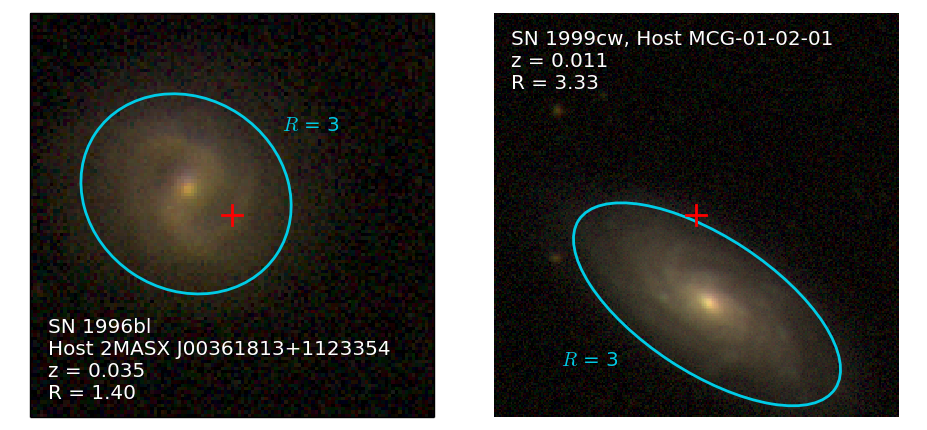}
\caption{SDSS $gri$ images of two spiral galaxies from our sample with SN positions marked
  in red and SExtractor-based isophotal radius estimates ($R = 3$) shown in
  blue.  We corrected SN 1996bl for dust but did not correct SN
  1999cw, as it exploded just outside the isophotal radius of its host
  galaxy and thus is beyond nearly all of its host galaxy's dust.}
\label{fig:rpar}
\end{figure}


There are three principal differences between our local dust correction and that of
R15.  First, for galaxies without star formation
rates (SFRs) from \citet{Neill09} (45\% of our sample), R15 used
morphological information to determine whether or not a galaxy was
globally star-forming.  However, GALEX NUV - SDSS $r$ magnitude is a more
reliable discriminator between passive and star-forming galaxies
(e.g. \citealp{Salim07}, their Fig. 1).  Passive galaxies have
NUV-$r$ $\gtrsim$ 5, while star-forming galaxies have NUV-$r$
$\lesssim$ 4.  For the 45\% of our sample with SDSS images, we
corrected for dust in galaxies that had NUV-$r$ $<$ 4.5 based on SExtractor
photometry \citep{Bertin96}.  For the final 19\% of our sample without
\citet{Neill09} SFR or SDSS images, we used morphology as an estimate of global star formation and
performed a local dust correction for Sa and later-type galaxies.  We
removed 3 morphologically ambiguous hosts from our sample (SN 2005eu,
SN 2006ah, and SN 2006is).


Second, SNe\,Ia near the edges of galaxies should have negligible
local dust.  We used SDSS and, when necessary, Digitized Sky Survey
images\footnote{\url{http://archive.eso.org/dss/dss}} to estimate the \citet{Sullivan06} SExtractor-based R
parameter, which gives the SN separation from the host normalized by the
size of the host galaxy.  For the 28\% of SNe approximately outside the isophotal
radius of their host galaxy (R $>$ 3; \citealp{Sullivan06}), we did
not correct for local dust regardless of the \citet{Salim07}
extinction estimate, which does not apply for passive, low-dust regions.  R15
dust-corrected all SNe in globally star-forming hosts, regardless of
the location of the SN.  Figure \ref{fig:rpar} shows two examples of
spiral host galaxies and their approximate isophotal radii.

In total,
our decision to apply or not to apply a dust correction was different from
that of R15 for 14\% of H09 SNe (13/92 SNe).  For 7 of these 13 SNe,
we did not apply a dust correction because the SN was outside the isophotal radius of its
host.  The other 6 SNe had morphology-based SF classifications that
disagreed with our NUV-$r$ data.

Finally, we adopted a slightly more conservative inclination cut, removing
galaxies with inclinations $>$70$^{\circ}$ based on the
\citet{Tully77} axial ratio method.  This removes an additional 16 SNe
from the JLA+PS1
sample and 11 from the R11 sample.  In total, 
the inclination cut removes $\sim$13\% of our sample.


\subsection{Varying the Baseline Analysis}
\label{sec:analysissyserr}

For a robust result, we performed several plausible variants of our baseline
analysis (R15 used a similar method to evaluate the
robustness of the LSF step).  We 
used the standard deviation of the measured LSF step
from all variations to estimate our systematic error.

Our FUV$-$NUV color measurements have a median signal-to-noise ratio
of 3.02.  Due to such large photometric uncertainties, the dust
correction and resulting $\Sigma_{\textrm{SFR}}$ is heavily affected by the 2 mag $A_{FUV}$
prior (e.g. SN 2003ic in Figure \ref{fig:galexlocal}).  Because
using this prior to correct for dust local to the SN\,Ia can have up to a $\sim$1 dex effect
on the measured $\Sigma_{\textrm{SFR}}$, we examined the effect of changing the Bayesian dust
prior to $A_{FUV} = 1.0\pm0.6$ and $A_{FUV} = 3.0\pm0.6$.  These values span
the full range of $A_{FUV}$ in blue galaxies measured by
\citet[see their
Figure 13]{Salim07}.  Changing this prior serves as a way to alleviate some of the uncertainty
associated with our global SFR determination; lowering this prior by 1
mag changes $\sim$10 SNe in our sample from Ia$\alpha$ to
Ia$\epsilon$.

Following R15, we tried an additional 3 local aperture
diameters between 2 and 6 kpc because the choice of a 4 kpc aperture is somewhat
arbitrary and other reasonable choices exist.  In part, the FWHM of
the FUV PSF determines the minimum spatial scale we can probe with
GALEX, which is approximately 2 kpc at our median redshift.  However,
Figure \ref{fig:galexlocal} shows that it is
still possible that a local aperture will encompass
components of a galaxy with different star-forming environments.  The
higher-resolution star formation maps of M33 in \citet{Boquien15}
show large $\Sigma_{\textrm{SFR}}$ variation on
much smaller, sub-kpc scales.  Nevertheless, we might hope that star-formation within a $\sim$few kpc
aperture is still much better correlated with the SN progenitor
environment than a global measurement due to the significant fraction of
prompt progenitors and low velocity dispersions of young stars \citep{deZeeuw99}.

The boundary between SNe\,Ia$\alpha$ and Ia$\epsilon$ is also
somewhat arbitrary.  We used values of \SSFR\ between  -3.1 and
-2.7.  For direct comparison
to \citet{Kelly15}, we also examined the boundary between star-forming
and passive of log($\Sigma_{SFR}$) $=-1.7$ and -1.85 (accounting for a
$\sim$0.4 dex offset between our SFR measurements and
\citealp{Kelly15}) when discussing Hubble
residual dispersion.

Finally, we tried using global rather than local star formation
(global star formation is a
less noisy measurement), and with or without 2.5$\sigma$-clipping.  Our list of
analysis variations is given in \S4, Table \ref{table:syserrtwo}.  

\begin{figure*}
\includegraphics[angle=0,width=7.in]{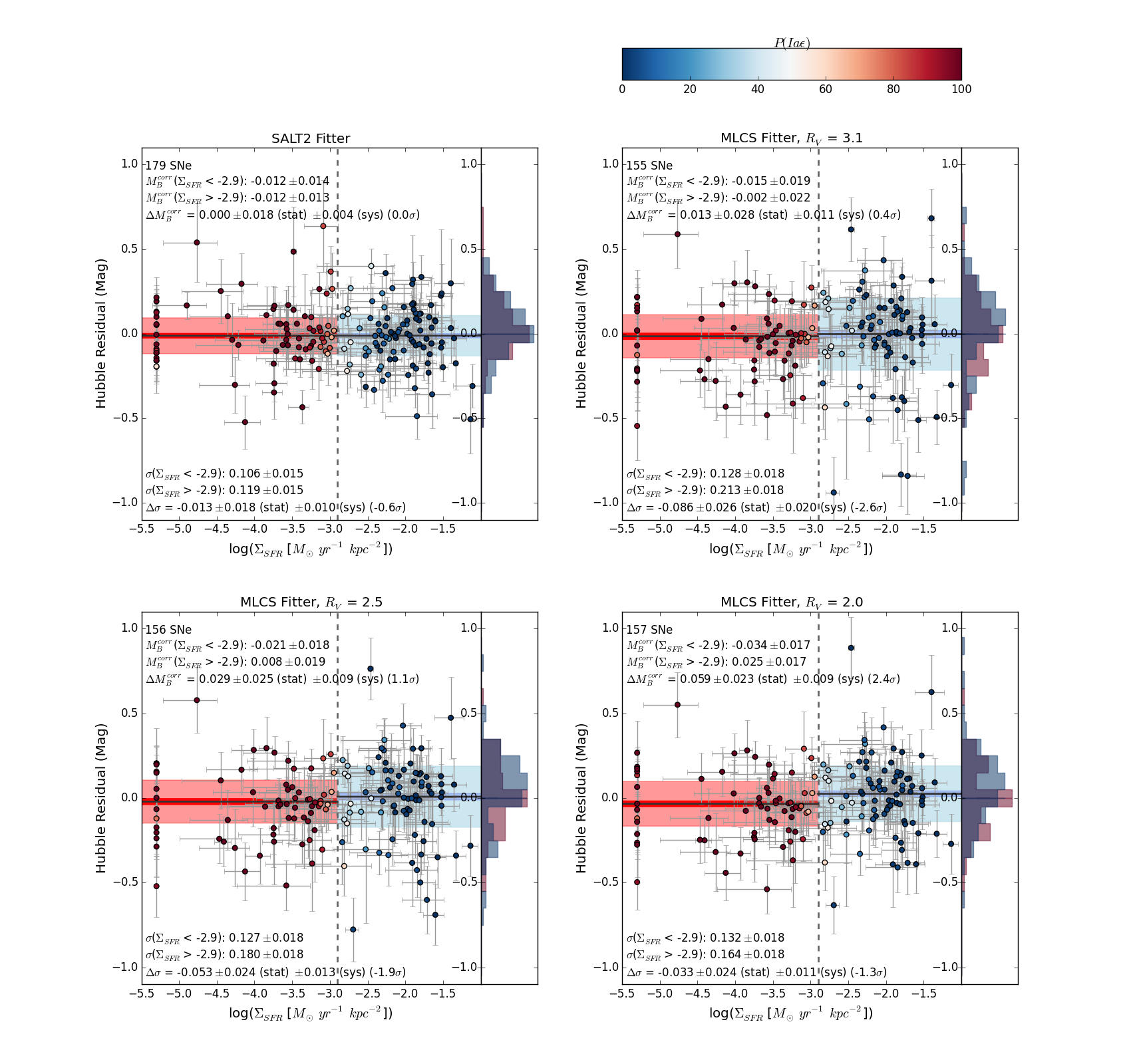}
\caption{Our baseline analysis for the JLA+PS1 sample (SALT2; upper left),
  and the R11 sample with different values of R$_V$ (MLCS2k2 fitter).
The color of each SN indicates the probability that it has a locally
passive environment, P(Ia$\epsilon$).  Shaded bars indicate the uncertainty on the mean (dark
shading; statistical error only) and the standard deviation of the maximum likelihood gaussian
(the weighted dispersion; light shading).  The LSF step is much smaller and has lower significance than the step found by R15,
although we detect it at \Rvlowbiassig$\sigma$ for the R$_V = 2.0$
case (\Rvlowbiassyssig$\sigma$ with systematic errors).  For R$_V =
3.1$ and 2.5, we find lower dispersion among SNe in locally passive
environments than those in locally star-forming environments at
3.3$\sigma$ and 2.2$\sigma$, respectively.   For consistency
with R15, SNe with only $\Sigma_{\textrm{SFR}}$ upper limits are
placed at \SSFR\ $=$ -5.3.  Systematic uncertainties are estimated
from several variants of our analysis (Table \ref{table:syserrtwo}).}
\label{fig:fullbias}
\end{figure*}

\section{Results}
\label{section:results}

We used \numJLA\
GALEX-detected SNe from JLA+PS1 and \numRiess\ SNe from R11 to measure
the LSF step and distance dispersion.  Although for certain variants of
the analysis, we see differences between SNe\,Ia$\epsilon$ and
Ia$\alpha$ at the level of $\sim$1-3$\sigma$, the evidence
for the LSF step is generally weak.

Although certain peculiar SNe
(e.g. SN 1991bg-like and SN 1991T-like) are not explicitly identified
and removed from these samples, the shape and color cuts applied by
JLA and R11 are sufficient to remove many of them.  However, we make
no effort to exclude peculiar SNe that JLA/R11 have determined to be
cosmologically useful so that we can directly 
assess the affect of local SF on the JLA/R11 cosmological analyses.
In contrast, \citet{Rigault13} and R15 remove identified SN 1991T
explicitly ($\sim$3\% of their sample).


In this section, we do not examine
the effect of correcting for the relationship between host mass and SN
distance \citep{Sullivan10} on the LSF step as only
$\sim$15\% of our SNe are low-mass hosts (log(M$_{\odot}$) $<$ 10;
R15 similarly found that few H09 SNe are in low-mass hosts).
However, we briefly consider its effect on H$_0$ in \S\ref{sec:H0}.  A
complete table with our GALEX measurements and Hubble residuals is
available online\footnote{\url{http://www.pha.jhu.edu/~djones/lsfstep.html}}, with
the first 25 rows given in Table \ref{table:measurements}.


\subsection{The Local Star Formation Step}
\label{sec:lsfbias}

We find a greatly reduced LSF step compared to R15 for all light curve
fitters and values of R$_V$.  Using SALT2, we find an LSF step of
\SALTbias$\pm$\SALTbiaserrfull\ mag.  With MLCS R$_V = 2.5$ (the value used
in the R11 baseline analysis), we find \Rvmidbias$\pm$\Rvmidbiaserrfull\ mag.
However, we do find mild evidence for an offset of
\Rvlowbias$\pm$\Rvlowbiaserrfull\ mag with R$_V = 2.0$ (\Rvlowbiassyssig$\sigma$
significance).  For R$_V = 3.1$, we found a value of
\Rvhighbias$\pm$\Rvhighbiaserrfull\ mag.
Our error budget includes systematic errors, which we estimated by measuring the
standard deviation of several variants of our analysis.  


Figure \ref{fig:fullbias} presents our baseline measurement of the LSF step and Hubble residual dispersion
for SNe\,Ia in locally passive and locally star-forming environments
(SNe\,Ia$\epsilon$ and SNe\,Ia$\alpha$, respectively), with colors indicating the probability
incorporated in our likelihood model that a
given SN\,Ia has a locally passive environment, P(Ia$\epsilon$).
We find that 47.2\% of R11 SNe in our sample are Ia$\epsilon$
and 46.0\% of JLA+PS1 SNe in our sample are Ia$\epsilon$.  The overall
intrinsic dispersion for our full MLCS sample ($\sim$0.13-0.17 mag;
0.14 for R$_V = 2.5$) is higher than for SALT2
(0.12 mag), likely due to the lack of recent calibration of MLCS2k2.
Intrinsic dispersion can also be affected by the distribution of light curve
parameters in the sample and the robustness of the photometric measurements. 

We find no significant difference in dispersion between
SNe\,Ia$\alpha$ and SNe\,Ia$\epsilon$ in SALT2.  In the R11 MLCS
sample, however, we find some evidence that SNe\,Ia$\epsilon$ have
lower dispersion ($\sigma_{Ia\epsilon}$) than SNe\,Ia$\alpha$.  For R$_V = 3.1$, the LSF
step is the lowest and $\sigma_{Ia\alpha}$ is the
highest (0.09 mag $>$ $\sigma_{Ia\epsilon}$; 2.6$\sigma$ with sys. error).
These results disagree with R15 at the 3$\sigma$ level.  For MLCS
with R$_V = 2.5$, $\sigma_{Ia\epsilon}$ is $\sim$0.05 mag less
than $\sigma_{Ia\alpha}$ (1.9$\sigma$ significance).  For R$_V
= 2.0$ we detected only a $\sim$0.03 mag difference in dispersion (1.3$\sigma$).  Our full results
for each analysis variant are presented in Table \ref{table:scatter}.

We found
that if we restrict to $z > 0.023$ (the R15 minimum $z$), we see more evidence for the LSF
step.  After this cut, there are 135 SALT2 SNe\,Ia and 104 MLCS SNe\,Ia.  The
increased significance of these results is expected because $\sim$3/4 of
our MLCS sample is from R15 when we apply this redshift cut.  For
MLCS R$_V = 2.0$, 2.5, and 3.1 we find LSF steps of 0.086$\pm$0.028
(3.1$\sigma$), 0.076$\pm$0.030 ($\sim$50\% of R15; 2.5$\sigma$), and 0.064$\pm$0.037
(35\% of R15; 1.8$\sigma$).  For SALT2, we only find a very small offset,
0.017$\pm$0.019 (18\% of the R15 result) at 0.9$\sigma$ significance.
The MLCS LSF steps are $\sim$50\% of those found by R15.  Except in
the case of MLCS with R$_V = 2.0$, the low-$z$
data alone ($0.01<z<0.023$) show slightly brighter SNe Ia$\alpha$ 
by $\sim$0.02-0.03 mag but with only 0.5$\sigma$ significance for
MLCS (0.06 mag with 1.4$\sigma$ for SALT2).  This effect is mostly due to $\sim$5
bright low-$z$ SNe, which do not have a large effect on the final
result (see the 2.5$\sigma$-clipping in Table \ref{table:syserrtwo}).  If the peculiar velocity corrections and
uncertainties for low-$z$ SNe were in error, we would expect, but do not observe, a significant increase
in uncertainty-weighted $M_B^{corr}$ dispersion below $z = 0.023$ (we
see $\lesssim$0.015 mag difference).  We did not find evidence that our highest-$z$ data ($z > 0.07$) were having a
significant effect on our results.

\subsection{Systematic Uncertainties}
\label{sec:syserr}

\begin{figure}
\includegraphics[angle=0,width=3.4in]{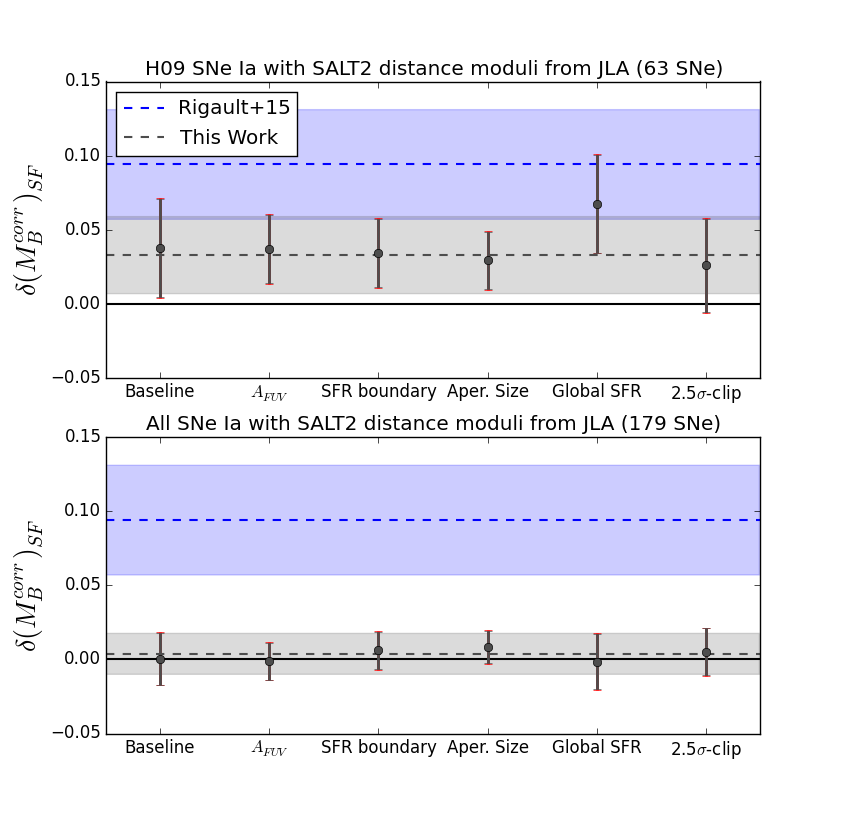}
\caption{The systematic error of the SALT2 LSF step estimated by
the effect of different variants of our analysis on the measurement of the
  LSF step.  Red error bars represent the standard deviation of all
  variants of our analysis added in quadrature to the uncertainties
  from each individual variant.  The top panel shows only H09 SNe included in
  \citet{Betoule14}, and the bottom panel shows our full SN\,Ia
  dataset.  The step we detect is $\sim$0.05 mag (1.3$\sigma$) with H09 SNe, but
  shrinks to $<$0.01 when we add in our full SN\,Ia sample.  The blue
  dashed lines and shaded regions show the R15 LSF step and 1$\sigma$
  uncertainty for SALT2.  The results from different variants of our
  analysis are very consistent; our measured systematic
  errors are only a small fraction of our statistical errors.  The global SFR variant is excluded
  from the systematic error calculation, as this is not a local measurement.}
\label{fig:syserr}
\end{figure}

\begin{figure}
\includegraphics[angle=0,width=3.4in]{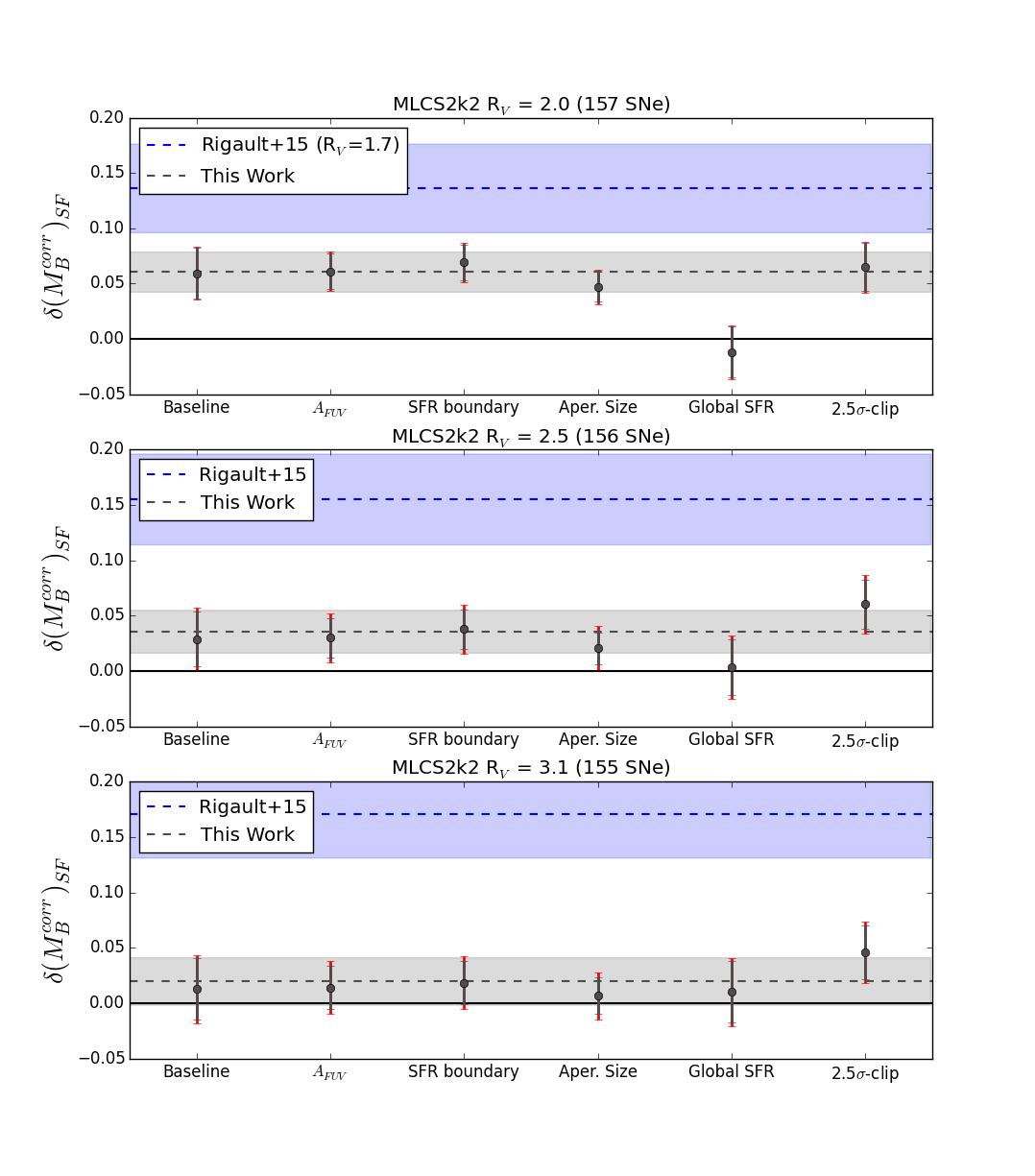}
\caption{The systematic error of the MLCS LSF step estimated by different variants of our analysis for R$_V = 2.0$, 2.5, 
and 3.1 in the R11 SN\,Ia sample.  The LSF step has \Rvlowbiassyssig$\sigma$ significance for
R$_V = 2.0$.  The baseline
analysis used to determine H$_0$ uses R$_V = 2.5$, for which we see
a small LSF step at \Rvmidbiassyssig$\sigma$ significance.  We see $<$1$\sigma$ significance
for R$_V = 3.1$.  The blue
  dashed lines and shaded regions show the R15 LSF step and 1$\sigma$
  uncertainties for MLCS2k2.  The global SFR variant is excluded
  from the systematic error calculation, as this is not a local measurement.}
\label{fig:syserrmlcs}
\end{figure}

Several different variants of our analysis
are consistent with the baseline
result.  The JLA+PS1 variants are shown visually in Figure
\ref{fig:syserr}, and the R11 variants are shown
in Figure \ref{fig:syserrmlcs}.  For the LSF step, the full results from both data sets are
presented in Table \ref{table:syserrtwo} and our dispersion results are 
presented in Table \ref{table:scatter}.  We have added the
standard deviation of the LSF step from all variants in
quadrature to our measured values (giving each type of variant,
e.g. aperture size, SFR boundary, etc., equal weight).  Because using
the global SFR is not truly a local measurement, we have excluded it
from our error computation but include it in our list of variants for comparison.

For nearly all samples, our most significant detections of the LSF step were
at a \SSFR\ boundary of -3.1 and a 3 kpc aperture radius.  For a
\SSFR\ boundary of -3.1, with SALT2 and MLCS R$_V = 2.5$ (the most relevant versions for cosmology), we
detected steps of 0.023$\pm$0.019 and 0.044$\pm$0.029, respectively.
These are $\sim$25\% of R15 values and insignificant.

For MLCS with R$_V = 2.5$ and 3.1, our
most significant detections came from the variant with
2.5$\sigma$-clipping.  They had
values of 0.060$\pm$0.026 mag (2.3$\sigma$) for R$_V = 2.5$ and
0.046$\pm$0.028 (1.6$\sigma$) for R$_V = 3.1$.  This may mean that
outliers are affecting our measurement.  However, we also expect
that they affect the R11 H$_0$ measurement in the same
way, and note that R$_V = 2.0$ 2.5$\sigma$-clipping has no significant effect.  

The variant with the smallest LSF step was the one based only upon
global SFR instead of local.  However, the significance of the difference 
is only $\lesssim$1$\sigma$ except in the case of R$_V = 2.0$.  The 
difference may stem from the fact that 25\% of SNe with globally
star-forming environments in our samples  had locally passive environments
(P(Ia$\epsilon$) $>$ 50\%).  Only 5\% of SNe with globally passive
environments had a $>$50\% probability of being locally SF.
Qualitatively, this agrees with H$\alpha$ data from \citet[their
Figure 5]{Rigault13}, who found that globally star-forming hosts often
had locally passive regions.



Even after adding the systematic error in quadrature, the MLCS R$_V =
2.0$ LSF step is detected at \Rvlowbiassyssig$\sigma$
(\Rvlowbias$\pm$\Rvlowbiaserrfull\ mag).  Future cosmology analyses using MLCS with
low R$_V$ should measure the LSF step in their samples to evaluate its
effect on cosmology.



The difference in the dispersion between the two SN populations in MLCS is
greatest in those same analysis variants discussed above, but as with our baseline
analysis, we see the opposite effect that R15 found.  We don't detect any
difference in dispersion for SALT2 with the exception of using global
instead of local SFR, for which we find a 0.05$\pm$0.018 mag (2.8$\sigma$)
reduction in dispersion for passive hosts.  For MLCS R$_V = 2.5$ and 3.1, we find a reduction in dispersion for locally passive SNe of
$\sim$0.05$-$0.1 mag ($\sim$1-3$\sigma$) for a \SSFR\ boundary of -3.1 and a 3
kpc aperture radius.

\subsection{Consistency with R15}
\label{sec:r15results}

\begin{deluxetable*}{p{0.3in}p{0.3in}p{0.3in}p{0.5in}cccccp{0.01in}ccccc}
\tablecaption{The Effect of Step-by-Step Changes in R15 Data,
  Distances, SFR Measurements, and Sample Cuts}
\renewcommand{\arraystretch}{1}
\tablehead{
\multicolumn{3}{c}{Measurements}&&
\multicolumn{5}{c}{SALT2} &&
\multicolumn{5}{c}{MLCS R$_V$=2.5}\\ 
\cline{1-3} \cline{5-9} \cline{11-15}\\
SN&&&SN&&&&&&&&&&&\\
sample&$\mu_{resid}$&$\Sigma_{\textrm{SFR}}$&
cuts&SNe&$\delta(M^{\textrm{corr}}_{B})_{SF}$\tablenotemark{a}&Sig.&
$\sigma_{Ia\alpha}$ $-$ $\sigma_{Ia\epsilon}$\tablenotemark{b}&Sig.&&
SNe&$\delta(M^{\textrm{corr}}_{B})_{SF}$\tablenotemark{a}&Sig.& 
$\sigma_{Ia\alpha}$ $-$ $\sigma_{Ia\epsilon}$\tablenotemark{b}& Sig.\\}
\startdata

H09&H09&R15&H09&77&0.093$\pm$0.026&3.5$\sigma$&-0.034$\pm$0.073&-0.5$\sigma$&&81&0.169$\pm$0.026&6.5$\sigma$&0.057$\pm$0.033&1.7$\sigma$\\*[2pt]
H09&H09&R15&\textbf{JPR}\tablenotemark{c},H09&59&0.129$\pm$0.030&4.3$\sigma$&0.012$\pm$0.047&0.2$\sigma$&&74&0.144$\pm$0.025&5.6$\sigma$&0.038$\pm$0.034&1.1$\sigma$\\*[2pt]
H09&\textbf{JPR}&R15&JPR,H09&59&0.062$\pm$0.032&1.9$\sigma$&0.030$\pm$0.031&1.0$\sigma$&&74&0.149$\pm$0.025&5.9$\sigma$&0.023$\pm$0.031&0.7$\sigma$\\*[2pt]
H09&JPR&\textbf{Here}\tablenotemark{d}&JPR,H09&59&0.071$\pm$0.033&2.2$\sigma$&0.009$\pm$0.031&0.3$\sigma$&&74&0.119$\pm$0.026&4.5$\sigma$&-0.010$\pm$0.030&-0.3$\sigma$\\*[2pt]
H09&JPR&Here&\textbf{JPR}&63&0.045$\pm$0.033&1.3$\sigma$&0.015$\pm$0.030&0.5$\sigma$&&78&0.097$\pm$0.027&3.6$\sigma$&-0.029$\pm$0.030&-1.0$\sigma$\\*[2pt]
JPR\tablenotemark{e}&JPR&Here&JPR,\textbf{\textit{z} $>$ 0.023}&135&0.017$\pm$0.019&0.9$\sigma$&-0.020$\pm$0.019&-1.1$\sigma$&&103&0.076$\pm$0.029&2.6$\sigma$&-0.041$\pm$0.029&-1.4$\sigma$\\*[2pt]
\textbf{JPR}\tablenotemark{e}&JPR&Here&JPR&179&0.000$\pm$0.018&0.0$\sigma$&-0.013$\pm$0.018&-0.7$\sigma$&&156&0.029$\pm$0.025&1.2$\sigma$&-0.053$\pm$0.024&-2.2$\sigma$\\*[2pt]
\enddata
\label{table:rigaultcomp}
\tablenotetext{a}{$\delta(M^{\textrm{corr}}_{B})_{SF}$ denotes the magnitude of the LSF step.}
\tablenotetext{b}{The difference in uncertainty-weighted dispersion
  between SNe\,Ia$\epsilon$ and Ia$\alpha$ (using the standard
  deviation of the maximum likelihood gaussians; $\sigma_{\epsilon}$
  and $\sigma_{\alpha}$ in Equation \ref{equation:li}).}
\tablenotetext{c}{\textbf{JLA+PS1} $M_B^{\textrm{corr}}$ for SALT2, \textbf{R11}
  $M_B^{\textrm{corr}}$ for R$_V = 2.5$.}
\tablenotetext{d}{Measurements of $\Sigma_{\textrm{SFR}}$ from
  \textbf{this work} (see \S3).}
\tablenotetext{e}{The full \textbf{JLA+PS1} (SALT2) and \textbf{R11} (MLCS) SN samples.}
\tablecomments{We show the difference between our analysis and R15 by
  improving one element of the analysis at a time.  We start with the
  R15 results and sequentially show the effect of adding light curve
  cuts from JLA+PS1/R11, using JLA/R11 distance moduli, using our updated
  SFR measurements, using only
  JLA/R11 (not H09) light curve cuts, and finally adding in the
  full SN samples with and without the R15 redshift cut of $z > 0.023$.  \textbf{The biggest differences come from
  adding the full sample for both SALT2 and MLCS and using improved SALT2
  distance moduli.}  The R11 SN light curve cuts
  also make a 1$\sigma$ difference in the MLCS results.  For
  consistency, we have used the likelihood minimizer used in the rest 
of this study to reproduce the R15 results (The SciPy Optimize package). This
minimizer returns smaller uncertainties than Minuit, which was used in
R15, but we find negligible differences in the maximum likelihood
values themselves.  The difference in LSF step we find for R15 data with
MLCS (our value is 0.014 mag higher) is because we adopt two separate dispersions
for SNe\,Ia$\alpha$ and SNe\,Ia$\epsilon$ whereas R15 use a single
value for the full sample.}
\end{deluxetable*}

R15 measured a much larger LSF step of 0.094$\pm$0.037
with SALT2, 0.155$\pm$0.041 with MLCS2k2 R$_V = 2.5$ and
0.171$\pm$0.040 with MLCS2k2 R$_V = 3.1$.  We did not directly compare
to their R$_V = 1.7$ data, but our R$_V = 2.0$ offset is 50\%
smaller than theirs.  Our
measured SALT2 LSF step has a \SALTdiscrep$\sigma$ discrepancy with the
R15 measurement, our MLCS2k2 R$_V = 2.5$ LSF step has a \Rvmiddiscrep$\sigma$
discrepancy, and our MLCS2k2 R$_V = 3.1$ LSF step has a
\Rvhighdiscrep$\sigma$ discrepancy.  



Table \ref{table:rigaultcomp} demonstrates the step by step impact of changes in the
R15 analysis or data, showing the effects of using the JLA+PS1 and R11 light curve
cuts, the JLA+PS1 and R11 distance moduli (with an updated
SALT2 light curve fitter for JLA+PS1), our improved \SSFR\
measurements, and using a larger
SN\,Ia sample (with and without the R15 $z > 0.023$ cut).


Updated distance moduli greatly decrease the significance of
the LSF step in JLA+PS1 data in SALT2, a 50\% reduction (a change in
significance of 2.4$\sigma$).  The
version of SALT2 used in recent analyses has an improved
SN\,Ia model and uncertainty propagation, a larger training sample,
and an updated value for $\beta$.  R11 distances
are nearly identical to H09 distances, so using these has no
significant effect on the LSF step.

Using our $\Sigma_{\textrm{SFR}}$ measurements increases the significance of the LSF
step by 0.3$\sigma$ for SALT2 and reduces it by 1.4$\sigma$
($\sim$20\%) for MLCS.
Between our data and the R15 data, there
is significant scatter in probability for 10\% $<$ P(Ia$\epsilon$) $<$
90\%, in large part due to our modest changes in dust correction
methodology.  However, 
we find only 3\% median offset in P(Ia$\epsilon$) between our data and
R15 and in \S\ref{sec:checks} we find that our method of $\Sigma_{\textrm{SFR}}$
measurement has little impact on the final results.  Our full set of $\Sigma_{\textrm{SFR}}$ measurements can be
compared to R15 using the data
we provide online and in Table \ref{table:measurements}.



There are 4 SNe in R11 and 4
SNe in JLA that pass R11/JLA light curve cuts but \textit{do not} pass H09 cuts (SNe
1992j, 1993h, 1999aw, 2001ic, 2006bd, 2006gt, 2007ba, and 2007cg).  We 
found that including them reduces the SALT2 LSF step by
a significant 37\% (0.9$\sigma$) and reduces the MLCS LSF step by
$\sim$15\% (0.9$\sigma$).  When applying any LSF-dependent effect to
cosmology, it is appropriate to match the cuts used in the
cosmological analysis to those used in the measurement.

For both the LSF step and the dispersion in MLCS, there is a
$>$1$\sigma$ change when we use the full SN\,Ia sample.  Although the
total statistical change from 3.6$\sigma$ to 1.2$\sigma$ is large, we do not expect this to be a result of peculiar velocity bias
from our low-$z$ data.
Some of the change may result from a greater sample dispersion, which reduces the
significance of small offsets.  A dispersion term is
typically added in quadrature to distance modulus uncertainties in cosmological
analyses, including R11 and \citet{Betoule14}, and has the same
effect. In addition, Table \ref{table:rigaultcomp} does not incorporate systematic error, which
may have an impact; high-$z$ data effectively have a larger aperture size due to a
PSF width that is a greater fraction of the 4 kpc aperture diameter.
Figure \ref{fig:syserrmlcs} shows that aperture variations may have up
to a 1$\sigma$ effect on the measured LSF step, and to expand our
sample size we have preferentially added low-$z$ data with smaller
effective apertures ($0.01<z<0.023$).

Table \ref{table:rigaultcomp} shows that the MLCS increase in
Ia$\alpha$ dispersion is mostly caused by the addition of new SNe
rather than to our $\Sigma_{\textrm{SFR}}$ measurements or new
distance moduli.  The surveys that comprise our sample typically have 
larger dispersion than H09, which
reduces the significance of the H09 sample.  There are a number of
possible sources for increased dispersion of a SN\,Ia sample,
including underestimating photometric difference image uncertainties
near bright hosts and nightly or absolute photometric calibration
uncertainties \citep{Scolnic14b}.  For MLCS, R11 may also have higher
sample dispersion because they make no cut on the $\chi^2$ of
the MLCS light curve fits, while H09 remove SNe with
reduced $\chi^2 > 1.5$.

\subsection{The Effect of MLCS Sample Cuts}
\label{sec:mlcssample}
\begin{figure}
\includegraphics[angle=0,width=3.5in]{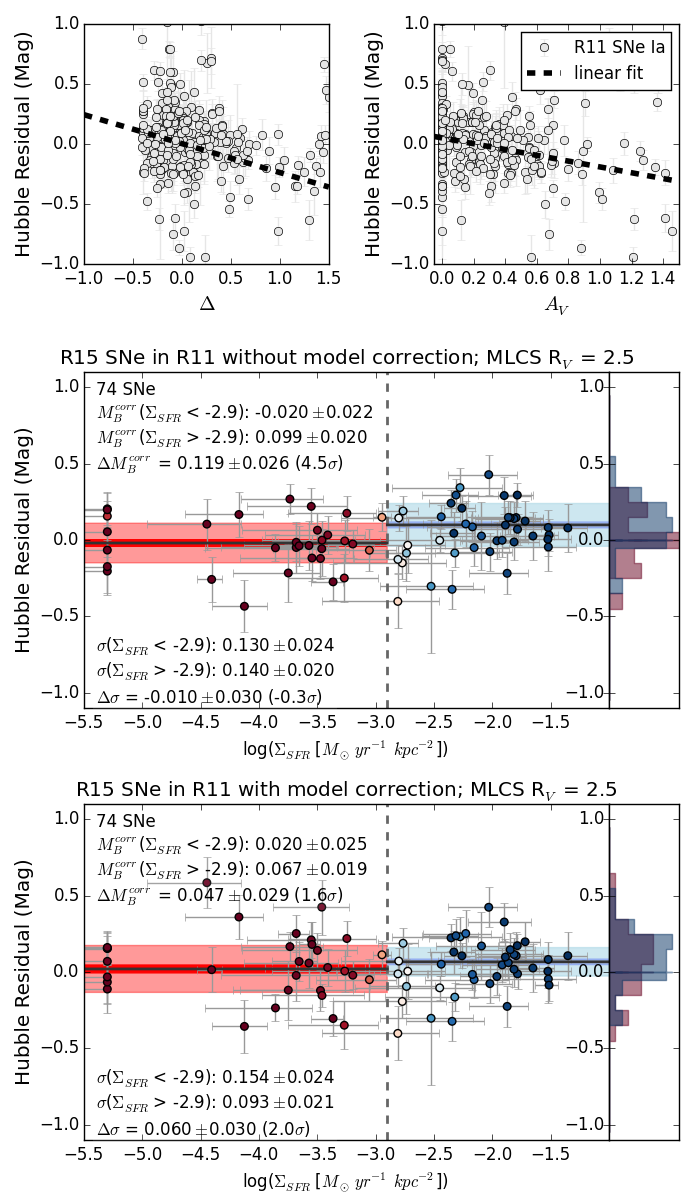}
\caption{A simple linear correction for Hubble residual trends in MLCS
reduces the significance of the R15 LSF step.  In the top panels, we
show MLCS $\Delta$ and $A_V$ fit to R11 SNe.  In the middle
panel, we show our measured SF bias using R11 SNe in H09.  In the
bottom panel, we make a linear correction for the MLCS Hubble residual trends, and the
LSF step is reduced from 4.5$\sigma$ to 1.6$\sigma$ significance.
Colors indicate P(Ia$\epsilon$), with P(Ia$\epsilon$)$\sim$100\% in
red and P(Ia$\epsilon$)$\sim$0 in blue.}
\label{fig:trends}
\end{figure}

In MLCS, the total difference of $\sim$0.14 mag between our analysis
and R15 may appear surprising, but in addition to the possible reasons
discussed above, much of the change between the R15 measurement and ours
appears to arise from the different demographics of the two samples and the
peculiarities of the MLCS light curve fitter.  H09 find
that for both high-A$_V$ SNe and high-$\Delta$ SNe, MLCS tends to
overcorrect leading to negative
residuals, and these negative residuals are not
subtle.  In our R$_V = 2.5$ sample, SNe with A$_V > $ 0.5 have a mean residual of
-0.22 mag, which has been seen elsewhere as evidence for a lower R$_V$
in high extinction environments.  Likewise, SNe with $\Delta > $ 0.7,
where the relation between light curve shape and luminosity becomes 
non-linear and is poorly sampled especially when MLCS2k2 was trained, 
have a mean residual of -0.23 mag.  Accordingly, the balance of rare
high A$_V$ SNe to rare high $\Delta$ SNe can affect an apparent LSF
step as the frequency of these objects correlates with host 
properties.

Passive hosts have preferentially higher $\Delta$ than SF
hosts (H09, their Figure 19), while SF hosts have preferentially higher
A$_V$.  In R15, the H09 data that have GALEX imaging
and pass their cuts contain
several SNe with large $\Delta$ but only two SNe with A$_V > 0.45$ for R$_V = 1.7$ (for R$_V = 3.1$, only two SNe with
A$_V > 0.7$).  Therefore a sample like R15 without high-A$_V$ hosts but \textit{with} high-$\Delta$ hosts
will have brighter passive SNe\,Ia on average, producing a larger apparent LSF step.


One approach to decrease sensitivity to MLCS Hubble
residual trends is to first remove the trends, and then determine the
LSF step.  In Figure \ref{fig:trends}, we fit a simple linear model to MLCS
Hubble residuals as a function of $\Delta$ and $A_V$, using R11 SNe in H09
(with $A_V<1.5$ and $\Delta<1.5$ to match H09).  When we correct for
those slopes, we see that the measured SF step using R11 SNe in H09
shrinks by a factor of 2.5 and is reduced from 4.5$\sigma$ to
1.6$\sigma$ significance.


SALT2 does not have the strong residual trends
with $X_1$ and $C$ that MLCS does with A$_V$ and $\Delta$, and we also find that restricting our sample to the H09
``best'' SALT2 cuts ($-0.1<C<0.2$) does not introduce an LSF step (but changing $\beta$ may;
see \S\ref{sec:dust}).  However, it is likely that recent substantial improvements
to the SALT2 model have removed some of the biases in its derived
distances.  Due to the lower dispersion of SALT2-fit SNe, the lack
of these residual trends, and because MLCS fits assume an extinction
law, it is likely that SALT2 is more effective at
standardizing SNe\,Ia.

In a future update of MLCS using a larger training sample, it would be
important to verify that these trends with host, A$_V$ and $\Delta$ are diminished.

\subsection{\citet{Kelly15} Scatter}

Using MLCS, \citet{Kelly15} see reduced
Hubble residual scatter of only 3.5\% in distance in highly star-forming regions (\SSFR\ $>$ -2.1 and
\SSFR\ $>$ -2.25).  Due to differences in
methodology, there is a $\sim$0.4 dex offset in $\Sigma_{SFR}$
measurements between our data and \citet{Kelly15}.  Because of this,
we adopt \SSFR\ $>$ -1.7 and \SSFR\ $>$ -1.85 as our $\Sigma_{SFR}$
boundaries for comparison.


In part, the low scatter seen by \citet{Kelly15} is because they
explicitly remove SNe with Hubble diagram residuals
$>$0.3 mag ($>$15\% in distance).  Because of this and because the R11
sample does not cut SNe with high extinction or large $\Delta$, our
unweighted standard deviation is a significantly larger
$\sim$0.25 (12\% in distance) for the R11 sample
at \SSFR\ $>$ -1.7 and \SSFR\ $>$ -1.85.  For SALT2,
the standard deviation is a slightly lower 0.20 mag, or 10\% in
distance, with no difference between SNe in locally passive/locally
star-forming environments.


We also see no difference in uncertainty-weighted dispersion for these
$\Sigma_{\textrm{SFR}}$ boundaries in SALT2, and we find that the
dispersion for SNe in both passive and star-forming environments in
SALT2 data is smaller than the \textit{lowest} dispersions we
observe with MLCS.  The scatter in our sample is much higher than in
\citet{Kelly15}, and we find a $\lesssim$0.02 mag ($\sim$14\%; $\sim$0.1-0.5$\sigma$)
reduction in dispersion for MLCS with R$_V = 2.0$.  SNe in star-forming
environments have \textit{higher} dispersion with low significance for
MLCS R$_V =2.5$.  For
R$_V = 3.1$, SNe in star-forming environments have $\sim$0.07 mag higher
dispersion at $\sim$1$\sigma$ significance.  A summary of 
our intrinsic dispersion measurements are in Table \ref{table:Kelly}.

If we apply H09 $\Delta$ and $A_V$ cuts to our data, we still see the
opposite effect as \citet{Kelly15}.  We can only reproduce the
\citet{Kelly15} results using their strict $\Delta$ and $A_V$ cuts,
which have not been used in any cosmological analysis to date.
However, these cuts may
prove useful in the future if this low-scatter population persists
when additional SNe are added to the data.

\subsection{Additional Consistency Checks}
\label{sec:checks}
We performed several consistency checks to verify that
individual SN datasets and differences between our analysis and R15
did not bias our results.  First,
we removed SNe discovered prior to the
year 2000, leaving 130 SNe
from JLA/PS1 and 116 SNe from R11.  Our results were
consistent with our baseline analysis; we measured a SALT2 LSF step of 0.010$\pm$0.025 mag and an MLCS
R$_V = 2.5$  step of 0.040$\pm$0.031 mag.  The R$_V = 2.0$ step was a
slightly higher, but consistent, 0.079$\pm$0.030 mag (2.7$\sigma$).  The dispersion of
SNe in highly SF regions was not significantly reduced.

Second, the photometry and calibration from low-$z$ surveys is not as
robust as recent data from SDSS and PS1.  The JLA/PS1 sample has 37
SNe with redshifts less than 0.1 that have GALEX data and pass our cuts, while the R11 sample includes no
SDSS/PS1 SNe as it predates them.  For comparison, we fit SDSS and PS1
SNe with MLCS to
see if the LSF step derived from these surveys alone are consistent with the
R11 results.  With SALT2, we find an LSF step of 0.034$\pm$0.028 mag
with lower SF dispersion by 0.049$\pm$0.024 mag (2.0$\sigma$).
With MLCS, we find a \textit{large} LSF step with 35 SNe of $\sim$0.14$\pm$0.055 mag with
1.7-2.9$\sigma$ significance.  As the sample consists of only $\sim$10-15
locally passive SNe, this step could still be caused by low statistics or
a limited range of light curve parameters comprising the sample.  As
discussed in \S\ref{sec:mlcssample}, the trends MLCS residuals have with different light curve parameters may
be a factor, as the size and significance of the LSF step is somewhat reduced when this sample is
restricted to low $\Delta$ and $A_V$.  This step
is also unlikely to affect recent cosmological analyses, which are
based on SALT2 or comprised mainly of low-$z$ data (e.g. R11, H09).
However, it is an interesting result that should be explored further
with photometric PS1 SNe and future DES data.
This sample is too small at \SSFR\ $>$ -1.85 for a reliable check on our
\citet{Kelly15} comparison.

If we make a host galaxy inclination cut at $>$80$^{\circ}$ following R15
(instead of our more conservative cut of $>$70$^{\circ}$), the results
are consistent with our baseline result, with MLCS LSF steps ranging from 0.00 mag
(R$_V = 3.1$) to 0.045 mag (R$_V = 2.0$) with uncertainties
$\sim$0.025 mag.  The SALT2 LSF step is
-0.016 mag ($<$1$\sigma$ significance).

Finally, we apply a dust correction to the FUV flux from
\textit{all} SN regions in star-forming hosts when determining
$\Sigma_{\textrm{SFR}}$, now including the 20 R11 SNe and 25 JLA/PS1 SNe with $R >
3$ (see \S\ref{sec:host ext}).  We again
find a comparable result; the SALT2 LSF step is 0.012$\pm$0.019 mag,
and the MLCS R$_V = 2.5$ LSF step is
0.040$\pm$0.028 mag.




\section{Discussion}

We find that local star formation has at little to no effect on
SN\,Ia distances in the R11 and JLA+PS1 samples.  
Our results have several important implications for cosmological analyses, H$_0$,
and future measurements of relationships between SNe\,Ia and their
host galaxy properties.

\subsection{The Effect of $\beta$ and R$_V$ on SN\,Ia Distances}
\label{sec:dust}
Although the modest differences we observe in mean magnitude and
dispersion for MLCS with certain values of R$_V$ could be due to the
relation between SN\,Ia progenitor properties
and derived distances, we consider it much more likely that host galaxy
extinction, which is highly correlated with star formation, 
is causing any observed bias.  We propose that some of the effects seen in R15,
\citet{Kelly15}, and our data may be due to dust rather than to a
secondary effect such as the progenitor age (e.g. \citealp{Childress14}).

With MLCS, the LSF step we found is 0.046$\pm$0.039 mag higher assuming R$_V =
2.0$ than assuming R$_V = 3.1$ (systematic errors added).  The
R$_V = 2.0$ LSF dispersion is 0.053$\pm$0.044 (stat+sys) mag lower than
R$_V = 3.1$.  It has been observed by several groups
(e.g. \citealp{Burns14}) that SNe\,Ia
in high-extinction environments have lower values of R$_V$.  Because
of this, it seems likely that the R$_V = 3.1$ extinction law is failing
to properly correct for the dust in some star-forming regions.


For SALT2, our value of $\beta$ has a value $\sim$0.6 higher in the latest
cosmological analyses than the value found in
H09.  This can have an important effect on the measured LSF step.  For
example, a SNe\,Ia in a locally star-forming
environment with $\sim$0.17 magnitudes of $A_V$,
would have its corrected magnitude shifted by 0.1 mag with this new
value of $\beta$.  For comparison, R15 SNe with locally star-forming 
environments have a mean fitted $A_V = 0.25$
for R$_V = 3.1$ and $A_V = 0.22$ for R$_V = 1.7$.  We don't see such
a large effect in our data, and would not expect $\beta$ to have the
exact effect of R$_V$, but
we do find that using a lowered $\beta$ of
2.5 (the value used in H09) in our analysis raises the SALT2 LSF step to 0.024$\pm$0.018 (1.3$\sigma$ significance).

In future cosmological analyses, it may be 
possible to separate star-forming and passive hosts and fit for 
two different values of $\beta$ or R$_V$.  This could
reduce scatter and provide more precise SN\,Ia distances for subsets
of the population, provided the systematic uncertainties in such an
analysis are well-understood.

The SALT2 light curve fitter shows the least difference
between SNe\,Ia$\epsilon$ and SNe\,Ia$\alpha$ $M_B^{corr}$ and also has the lowest
dispersion in both star-forming and passive regions.  The lowest dispersion
we find using MLCS is still higher than the SALT2 dispersion for both SNe
Ia$\epsilon$ and Ia$\alpha$.
For this reason, SALT2 may be a more reliable light curve fitter for
cosmological analyses.  In its current version, MLCS fails to standardize SNe\,Ia
to the extent that SALT2 does and has fitter biases that correlate with host
properties (such as Hubble residual nonlinearities with high $\Delta$
and an assumed value for R$_V$).  Perhaps a re-trained
version of MLCS that incorporates terms such as random SN color scatter
\citep{Scolnic14} would reduce the MLCS
outlier fraction and provide more precise distances.

\subsection{The Effect on Measuring H$_0$}
\label{sec:H0}

Because our final measurement of the LSF step with R$_V = 2.5$ is only
a \Rvmidbiassig$\sigma$ detection, there are no grounds in the
Bayesian sense to correct H$_0$
for the LSF step.  However, a useful test of systematic uncertainties in
the future will be to use only star-forming hosts in the Hubble flow
sample, which have similar physical properties to the nearby Cepheid-calibrated
sample and will better control for unknown biases in
metallicity, dust, or progenitor age.

Adopting the 47.2\% SN\,Ia$\epsilon$ fraction we find for R11 and the
7.0\% SN\,Ia$\epsilon$ fraction found by R15 for the Cepheid sample with Equation
\ref{eq:H0}, we find no evidence for a reduced value of H$_0$.
Following R15, if we were to replace the host mass step with the LSF
step, our measurement suggests a 0.1\% increase in H$_0$ because
the size of the LSF correction is slightly less than the size of the
host mass correction.


One caveat is that R11 added the MLCS intrinsic
SN\,Ia dispersion but not the full apparent intrinsic dispersion in quadrature to the distance modulus
uncertainties in their Hubble flow SNe.  We find
that forcing our maximum likelihood gaussian model to use only the MLCS 
intrinsic dispersion of 0.08 mag raises the magnitude of
the R$_V = 2.5$ LSF step we derive to 0.045$\pm$0.019 (a 2.4$\sigma$ detection,
but 2.1$\sigma$ with systematic uncertainty added).
This could be because it allows outliers to have a greater effect on
the measurement.  However, applying
this correction after removing the host mass step still only results
in a reduction in H$_0$ of 0.11 km s$^{-1}$ Mpc$^{-1}$.  The R11 value for
H$_0$ is within the 1$\sigma$ uncertainty of the LSF step.  The
highest LSF step we are able to find using all our analysis variants
with 0.08 mag dispersion is
0.066$\pm$0.22 mag (the 2.5$\sigma$-clipped variant), and even this extreme measurement lowers H$_0$ by only
0.4 km s$^{-1}$ Mpc$^{-1}$.

Finally, if we measure the LSF step \textit{after} host mass correction using
masses from \citet[53\% of the R11 sample]{Neill09} and again using a
dispersion of 0.08 mag, we find a LSF
step of 0.023$\pm$0.027 (stat+sys) mag for R$_V = 2.5$.  This results in a
small reduction of 0.3 km s$^{-1}$ Mpc$^{-1}$.  Because we
detect this effect at $<$1$\sigma$ (with systematic error added in
quadrature), we do not believe a correction is justified.


\subsection{Future Measurement of the  LSF Step}

Although we have only detected the LSF step at low
significance with GALEX FUV data, GALEX alone is not the best tool for
studying local regions due to its large PSF width and the uncertain UV
extinction correction.  The LSF step
would be best identified in local H$\alpha$ (e.g. \citealp{Rigault13}), high-resolution UV data
from the \textit{Hubble Space Telescope} (HST), or local SED fitting.  

Table \ref{table:rigaultcomp} shows that sample selection has a
significant effect on our results.  We suggest that studies examining host galaxy effects use
the same SN\,Ia samples and selection criteria as the latest cosmology
analyses when possible.  It may
be possible to detect the LSF step or differences in dispersion at
higher significance using different light curve or distance
modulus cuts, but the results of such
analyses would not necessarily apply to typical measurements of cosmological
parameters.

Local SED fitting may be the optimal approach for studying the
relation between host galaxy properties and SN\,Ia distances,
as it can put simultaneous (albeit sometimes degenerate) constraints 
on a number of parameters that may
correlate with SN\,Ia distances such as stellar age, extinction, star
formation history, and mass contained in a local region.  Approaches that don't depend
entirely on
GALEX data will also be able to measure local regions at higher redshifts
and put better contraints on possible redshift-dependent biases.

The size of the samples with which we can
examine the effects of host galaxy properties on SN\,Ia corrected
magnitudes will increase dramatically in the next few years.  The PS1 photometric sample 
alone will consist of up to $\sim$2,000 SNe\,Ia with
cosmologically-useful light curves.  The Dark Energy Survey (DES) will
contribute thousands more up to redshifts of $\sim$1.  Although
measurements of local regions become more difficult at high-$z$, a
ground-based optical survey with PSF FWHM $\sim$1 arcsec will be able
to use a much larger SN sample provided the absence of UV data is not prohibitive.  Surveys such as
PS1 or DES are able to examine local regions of 5 kpc
diameter, similar in size to the apertures used in this study,
up to $z \simeq 0.35$.

\section{Conclusions}

Analyzing the same SNe\,Ia used to determine the most recent values of $w$
and H$_0$, we find little evidence for a LSF step, which suggests that
correcting cosmological parameters for this effect is not necessary.  There is 
only \Rvmidbiassyssig$\sigma$ evidence for the LSF step in 
R11 MLCS data assuming R$_V = 2.5$ (the R$_V$ R11 used in their 
baseline analysis) and \SALTbiassyssig$\sigma$ evidence for the LSF step in JLA+PS1
SALT2 data.  Our most significant detection
uses MLCS data assuming R$_V = 2.0$, for which we find
\Rvlowbiassyssig$\sigma$ evidence for a step.
The sizes of both of these steps are greatly reduced compared to the
measurement of R15.  Lower values of $\beta$ in SALT2
and R$_V$ in MLCS may increase the size
and the significance of the LSF step.

Compared to R15, differences in our $\Sigma_{SFR}$ measurement and
dust correction technique reduced the size
of the MLCS LSF step by $\sim$20\% and increased the
SALT2 LSF step by $\sim$15\%.  Using MLCS
sample cuts from R11 reduced the offset by an additional
$\sim$20\% and adding the full R11 sample reduced the offset to
\Rvmidbias$\pm$\Rvmidbiaserrfull\ mag, likely due to the higher dispersion and better
statistics of the full
sample.  Using new distance moduli and sample cuts from only
\citet{Betoule14} (and not H09) reduced the SALT2 LSF step by 60\% and using the full
JLA+PS1 sample reduced the SALT2 step to a value of \SALTbias$\pm$\SALTbiaserrfull\ mag.

MLCS sample cuts have a significant impact on the results. 
MLCS Hubble diagram residuals are more negative at
greater A$_V$ and $\Delta$, which must be carefully taken into account
in cosmological analyses.  In particular, passive hosts are known to have
preferentially higher $\Delta$
but lower A$_V$ (H09).  We suspect that because the
R15 sample had few high-A$_V$ SNe but a wide range of $\Delta$, their
locally star-forming SNe had preferentially fainter Hubble residuals.

We found that JLA+PS1 SNe fit with SALT2 had lower dispersion than
MLCS-fit R11 SNe in star-forming \textit{or} passive environments.
We also found that locally star-forming SNe in our sample did not have lower
dispersion at \SSFR\ $>$ -2.9.  In MLCS with R$_V = 3.1$,
SNe\,Ia in locally passive environments have lower dispersion 
than those in locally star-forming environments by $\sim$0.09 mag, a
2.5$\sigma$ result.  Using MLCS with R$_V = 2.5$, we see
a 0.053$\pm$0.029 mag difference.

The lowest SN\,Ia dispersions come from using SALT2 distance
moduli.  In contrast to \citet{Kelly15}, with MLCS we found no 
evidence that SNe in highly star-forming environments 
have lower dispersion than locally passive SNe
using R$_V = 2.0$.  With R$_V = 3.1$ we found that SNe
in star-forming environments had \textit{greater} dispersion ($\sim$1-2$\sigma$ significance), but note that we did not make the
\citet{Kelly15} sample cuts.  We can only reproduce the
\citet{Kelly15} results by using their strict cuts on the SN light curve parameters
$\Delta$ and $A_V$ and removing SNe with Hubble residuals $>$0.3 mag,
which restricts our sample to largely the same data as \citet{Kelly15}. 




The LSF step may also be difficult to detect because
of the large PSF width of GALEX and it may also be that the LSF step
is only apparent in analyses with certain types of light curve
selection or outlier rejection.  Future studies with local
H$\alpha$, SED fitting, or HST UV observations will have an improved
ability to detect local effects.  Our results also show that certain
SN sample cuts may inadvertently increase biases in cosmology.  We expect that with the large SN\,Ia samples
from PS1 and DES that will be published in the next few years,
the systematic uncertainties on H$_0$ and the
dark energy equation of state will come into clearer focus.

\acknowledgements

This work would not have been possible without comments, suggestions,
and other assistance from Mickael Rigault.  We would also like to thank the 
anonymous referee and Pat Kelly for many useful comments
and suggestions.

\appendix
\section{Calculation of Probabilities and Maximum Likelihood Estimation}
\label{sec:maxlike}

The only significant difference between our method of measuring
the maximum likelihood LSF step and Hubble residual dispersions and
the R15 method is that
we allowed the intrinsic dispersion of both SN\,Ia populations
(Ia$\epsilon$ and Ia$\alpha$) to be fit
by our maximum likelihood model.  We
describe our full procedure below.

We first converted the dust-corrected FUV flux
into $\Sigma_{\textrm{SFR}}$ following R15 (their Equation 1).
We set the boundary between the locally star-forming and locally
passive population at log($\Sigma_{SFR}$) $ = -2.9$ as in R15, and measured the
probability that the SN\,Ia exploded in a locally passive environment
based on the full probability distribution from our dust-corrected
photometric measurements.


We used these probabilities to construct a maximum likelihood
model assuming two gaussian populations of SNe with different mean Hubble
residuals and dispersions.  The likelihood is determined by the
equation:
\begin{equation}
\begin{split}
\mathcal{L}_i = P(Ia\alpha) \times \frac{1}{\sqrt{2\pi(\sigma_i^2+\sigma_{\alpha}^2)}}\exp(-\frac{(M_{B,i}^{corr}-\mu_{\alpha})^2}{2(\sigma_i^2+\sigma_{\alpha}^2)})\\
 +  P(Ia\epsilon) \times \frac{1}{\sqrt{2\pi(\sigma_i^2+\sigma_{\epsilon}^2)}}\exp(-\frac{(M_{B,i}^{corr}-\mu_{\epsilon})^2}{2(\sigma_i^2+\sigma_{\epsilon}^2)}),
\end{split}
\label{equation:li}
\end{equation}

\noindent where $M_{B,i}^{corr}$ is the corrected magnitude
and $\sigma_i$ is the corrected magnitude
uncertainty of a given SN\,Ia.  P(Ia$\alpha$) and P(Ia$\epsilon$)
are the probabilities that the SN environment is locally star-forming
or locally passive, respectively.   $\mu_\alpha$, $\mu_\epsilon$, $\sigma_\alpha$ and
$\sigma_\epsilon$ are free parameters equal to the means and standard deviations of the
normal distributions of SNe\,Ia$\alpha$ and Ia$\epsilon$.  To determine what
these parameters are, we found the maximum likelihood model by minimizing:

\begin{equation}
\textrm{log}(\mathcal{L}) = -2\sum\limits_{i=1}^{N}\textrm{log}(\mathcal{L}_i)
\end{equation}

\noindent where $N$ is the number of SNe\,Ia in the sample.

Instead of adding an intrinsic dispersion term
in quadrature to the Hubble residuals such that the reduced $\chi^2$
of the sample is 1, as is commonly done in
cosmological analyses (and in R15), we fit to the
standard deviations of our gaussian maximum 
likelihood model for SNe\,Ia$\alpha$ and Ia$\epsilon$.  We
verified that allowing the dispersion to be fit by our model instead
of specifying it beforehand does not affect our results.




\begin{thebibliography}{51}
\expandafter\ifx\csname natexlab\endcsname\relax\def\natexlab#1{#1}\fi

\bibitem[{{Aldering} {et~al.}(2002){Aldering}, {Adam}, {Antilogus}, {Astier},
  {Bacon}, {Bongard}, {Bonnaud}, {Copin}, {Hardin}, {Henault}, {Howell},
  {Lemonnier}, {Levy}, {Loken}, {Nugent}, {Pain}, {Pecontal}, {Pecontal},
  {Perlmutter}, {Quimby}, {Schahmaneche}, {Smadja}, \&
  {Wood-Vasey}}]{Aldering02}
{Aldering}, G., {Adam}, G., {Antilogus}, P., {et~al.} 2002, in Society of
  Photo-Optical Instrumentation Engineers (SPIE) Conference Series, Vol. 4836,
  Survey and Other Telescope Technologies and Discoveries, ed. J.~A. {Tyson} \&
  S.~{Wolff}, 61--72

\bibitem[{{Bertin} \& {Arnouts}(1996)}]{Bertin96}
{Bertin}, E., \& {Arnouts}, S. 1996, \aaps, 117, 393

\bibitem[{{Betoule} {et~al.}(2014){Betoule}, {Kessler}, {Guy}, {Mosher},
  {Hardin}, {Biswas}, {Astier}, {El-Hage}, {Konig}, {Kuhlmann}, {Marriner},
  {Pain}, {Regnault}, {Balland}, {Bassett}, {Brown}, {Campbell}, {Carlberg},
  {Cellier-Holzem}, {Cinabro}, {Conley}, {D'Andrea}, {DePoy}, {Doi}, {Ellis},
  {Fabbro}, {Filippenko}, {Foley}, {Frieman}, {Fouchez}, {Galbany}, {Goobar},
  {Gupta}, {Hill}, {Hlozek}, {Hogan}, {Hook}, {Howell}, {Jha}, {Le Guillou},
  {Leloudas}, {Lidman}, {Marshall}, {M{\"o}ller}, {Mour{\~a}o}, {Neveu},
  {Nichol}, {Olmstead}, {Palanque-Delabrouille}, {Perlmutter}, {Prieto},
  {Pritchet}, {Richmond}, {Riess}, {Ruhlmann-Kleider}, {Sako}, {Schahmaneche},
  {Schneider}, {Smith}, {Sollerman}, {Sullivan}, {Walton}, \&
  {Wheeler}}]{Betoule14}
{Betoule}, M., {Kessler}, R., {Guy}, J., {et~al.} 2014, \aap, 568, A22

\bibitem[{{Boquien} {et~al.}(2015){Boquien}, {Calzetti}, {Aalto}, {Boselli},
  {Braine}, {Buat}, {Combes}, {Israel}, {Kramer}, {Lord}, {Relano},
  {Rosolowsky}, {Stacey}, {Tabatabaei}, {van der Tak}, {van der Werf},
  {Verley}, \& {Xilouris}}]{Boquien15}
{Boquien}, M., {Calzetti}, D., {Aalto}, S., {et~al.} 2015, ArXiv e-prints

\bibitem[{{Burns} {et~al.}(2014){Burns}, {Stritzinger}, {Phillips}, {Hsiao},
  {Contreras}, {Persson}, {Folatelli}, {Boldt}, {Campillay}, {Castell{\'o}n},
  {Freedman}, {Madore}, {Morrell}, {Salgado}, \& {Suntzeff}}]{Burns14}
{Burns}, C.~R., {Stritzinger}, M., {Phillips}, M.~M., {et~al.} 2014, \apj, 789,
  32

\bibitem[{{Cardelli} {et~al.}(1989){Cardelli}, {Clayton}, \&
  {Mathis}}]{Cardelli89}
{Cardelli}, J.~A., {Clayton}, G.~C., \& {Mathis}, J.~S. 1989, \apj, 345, 245

\bibitem[{{Childress} {et~al.}(2013){Childress}, {Aldering}, {Antilogus},
  {Aragon}, {Bailey}, {Baltay}, {Bongard}, {Buton}, {Canto}, {Cellier-Holzem},
  {Chotard}, {Copin}, {Fakhouri}, {Gangler}, {Guy}, {Hsiao}, {Kerschhaggl},
  {Kim}, {Kowalski}, {Loken}, {Nugent}, {Paech}, {Pain}, {Pecontal}, {Pereira},
  {Perlmutter}, {Rabinowitz}, {Rigault}, {Runge}, {Scalzo}, {Smadja}, {Tao},
  {Thomas}, {Weaver}, \& {Wu}}]{Childress13}
{Childress}, M., {Aldering}, G., {Antilogus}, P., {et~al.} 2013, \apj, 770, 108

\bibitem[{{Childress} {et~al.}(2014){Childress}, {Wolf}, \&
  {Zahid}}]{Childress14}
{Childress}, M.~J., {Wolf}, C., \& {Zahid}, H.~J. 2014, \mnras, 445, 1898

\bibitem[{{Conley} {et~al.}(2011){Conley}, {Guy}, {Sullivan}, {Regnault},
  {Astier}, {Balland}, {Basa}, {Carlberg}, {Fouchez}, {Hardin}, {Hook},
  {Howell}, {Pain}, {Palanque-Delabrouille}, {Perrett}, {Pritchet}, {Rich},
  {Ruhlmann-Kleider}, {Balam}, {Baumont}, {Ellis}, {Fabbro}, {Fakhouri},
  {Fourmanoit}, {Gonz{\'a}lez-Gait{\'a}n}, {Graham}, {Hudson}, {Hsiao},
  {Kronborg}, {Lidman}, {Mourao}, {Neill}, {Perlmutter}, {Ripoche}, {Suzuki},
  \& {Walker}}]{Conley11}
{Conley}, A., {Guy}, J., {Sullivan}, M., {et~al.} 2011, \apjs, 192, 1

\bibitem[{{de Zeeuw} {et~al.}(1999){de Zeeuw}, {Hoogerwerf}, {de Bruijne},
  {Brown}, \& {Blaauw}}]{deZeeuw99}
{de Zeeuw}, P.~T., {Hoogerwerf}, R., {de Bruijne}, J.~H.~J., {Brown}, A.~G.~A.,
  \& {Blaauw}, A. 1999, \aj, 117, 354

\bibitem[{{Dom{\'{\i}}nguez} {et~al.}(2001){Dom{\'{\i}}nguez}, {H{\"o}flich},
  \& {Straniero}}]{Dominguez01}
{Dom{\'{\i}}nguez}, I., {H{\"o}flich}, P., \& {Straniero}, O. 2001, \apj, 557,
  279

\bibitem[{{Ganeshalingam} {et~al.}(2010){Ganeshalingam}, {Li}, {Filippenko},
  {Anderson}, {Foster}, {Gates}, {Griffith}, {Grigsby}, {Joubert}, {Leja},
  {Lowe}, {Macomber}, {Pritchard}, {Thrasher}, \& {Winslow}}]{Ganeshalingam10}
{Ganeshalingam}, M., {Li}, W., {Filippenko}, A.~V., {et~al.} 2010, \apjs, 190,
  418

\bibitem[{{Guy} {et~al.}(2007){Guy}, {Astier}, {Baumont}, {Hardin}, {Pain},
  {Regnault}, {Basa}, {Carlberg}, {Conley}, {Fabbro}, {Fouchez}, {Hook},
  {Howell}, {Perrett}, {Pritchet}, {Rich}, {Sullivan}, {Antilogus}, {Aubourg},
  {Bazin}, {Bronder}, {Filiol}, {Palanque-Delabrouille}, {Ripoche}, \&
  {Ruhlmann-Kleider}}]{Guy07}
{Guy}, J., {Astier}, P., {Baumont}, S., {et~al.} 2007, \aap, 466, 11

\bibitem[{{Guy} {et~al.}(2010){Guy}, {Sullivan}, {Conley}, {Regnault},
  {Astier}, {Balland}, {Basa}, {Carlberg}, {Fouchez}, {Hardin}, {Hook},
  {Howell}, {Pain}, {Palanque-Delabrouille}, {Perrett}, {Pritchet}, {Rich},
  {Ruhlmann-Kleider}, {Balam}, {Baumont}, {Ellis}, {Fabbro}, {Fakhouri},
  {Fourmanoit}, {Gonz{\'a}lez-Gait{\'a}n}, {Graham}, {Hsiao}, {Kronborg},
  {Lidman}, {Mourao}, {Perlmutter}, {Ripoche}, {Suzuki}, \& {Walker}}]{Guy10}
{Guy}, J., {Sullivan}, M., {Conley}, A., {et~al.} 2010, \aap, 523, A7

\bibitem[{{Hamuy} {et~al.}(1996){Hamuy}, {Phillips}, {Suntzeff}, {Schommer},
  {Maza}, \& {Aviles}}]{Hamuy96}
{Hamuy}, M., {Phillips}, M.~M., {Suntzeff}, N.~B., {et~al.} 1996, \aj, 112,
  2398

\bibitem[{{Hamuy} {et~al.}(2006){Hamuy}, {Folatelli}, {Morrell}, {Phillips},
  {Suntzeff}, {Persson}, {Roth}, {Gonzalez}, {Krzeminski}, {Contreras},
  {Freedman}, {Murphy}, {Madore}, {Wyatt}, {Maza}, {Filippenko}, {Li}, \&
  {Pinto}}]{Hamuy06}
{Hamuy}, M., {Folatelli}, G., {Morrell}, N.~I., {et~al.} 2006, \pasp, 118, 2

\bibitem[{{Hayden} {et~al.}(2013){Hayden}, {Gupta}, {Garnavich}, {Mannucci},
  {Nichol}, \& {Sako}}]{Hayden13}
{Hayden}, B.~T., {Gupta}, R.~R., {Garnavich}, P.~M., {et~al.} 2013, \apj, 764,
  191

\bibitem[{{Hicken} {et~al.}(2009{\natexlab{a}}){Hicken}, {Wood-Vasey},
  {Blondin}, {Challis}, {Jha}, {Kelly}, {Rest}, \& {Kirshner}}]{Hicken09}
{Hicken}, M., {Wood-Vasey}, W.~M., {Blondin}, S., {et~al.} 2009{\natexlab{a}},
  \apj, 700, 1097

\bibitem[{{Hicken} {et~al.}(2009{\natexlab{b}}){Hicken}, {Challis}, {Jha},
  {Kirshner}, {Matheson}, {Modjaz}, {Rest}, {Wood-Vasey}, {Bakos}, {Barton},
  {Berlind}, {Bragg}, {Brice{\~n}o}, {Brown}, {Caldwell}, {Calkins}, {Cho},
  {Ciupik}, {Contreras}, {Dendy}, {Dosaj}, {Durham}, {Eriksen}, {Esquerdo},
  {Everett}, {Falco}, {Fernandez}, {Gaba}, {Garnavich}, {Graves}, {Green},
  {Groner}, {Hergenrother}, {Holman}, {Hradecky}, {Huchra}, {Hutchison},
  {Jerius}, {Jordan}, {Kilgard}, {Krauss}, {Luhman}, {Macri}, {Marrone},
  {McDowell}, {McIntosh}, {McNamara}, {Megeath}, {Mochejska}, {Munoz},
  {Muzerolle}, {Naranjo}, {Narayan}, {Pahre}, {Peters}, {Peterson}, {Rines},
  {Ripman}, {Roussanova}, {Schild}, {Sicilia-Aguilar}, {Sokoloski}, {Smalley},
  {Smith}, {Spahr}, {Stanek}, {Barmby}, {Blondin}, {Stubbs}, {Szentgyorgyi},
  {Torres}, {Vaz}, {Vikhlinin}, {Wang}, {Westover}, {Woods}, \&
  {Zhao}}]{Hicken09b}
{Hicken}, M., {Challis}, P., {Jha}, S., {et~al.} 2009{\natexlab{b}}, \apj, 700,
  331

\bibitem[{{Hicken} {et~al.}(2012){Hicken}, {Challis}, {Kirshner}, {Rest},
  {Cramer}, {Wood-Vasey}, {Bakos}, {Berlind}, {Brown}, {Caldwell}, {Calkins},
  {Currie}, {de Kleer}, {Esquerdo}, {Everett}, {Falco}, {Fernandez},
  {Friedman}, {Groner}, {Hartman}, {Holman}, {Hutchins}, {Keys}, {Kipping},
  {Latham}, {Marion}, {Narayan}, {Pahre}, {Pal}, {Peters}, {Perumpilly},
  {Ripman}, {Sipocz}, {Szentgyorgyi}, {Tang}, {Torres}, {Vaz}, {Wolk}, \&
  {Zezas}}]{Hicken12}
{Hicken}, M., {Challis}, P., {Kirshner}, R.~P., {et~al.} 2012, \apjs, 200, 12

\bibitem[{{Hsiao} {et~al.}(2007){Hsiao}, {Conley}, {Howell}, {Sullivan},
  {Pritchet}, {Carlberg}, {Nugent}, \& {Phillips}}]{Hsiao07}
{Hsiao}, E.~Y., {Conley}, A., {Howell}, D.~A., {et~al.} 2007, \apj, 663, 1187

\bibitem[{{Hudson} {et~al.}(2004){Hudson}, {Smith}, {Lucey}, \&
  {Branchini}}]{Hudson04}
{Hudson}, M.~J., {Smith}, R.~J., {Lucey}, J.~R., \& {Branchini}, E. 2004,
  \mnras, 352, 61

\bibitem[{{Jha} {et~al.}(2007){Jha}, {Riess}, \& {Kirshner}}]{Jha07}
{Jha}, S., {Riess}, A.~G., \& {Kirshner}, R.~P. 2007, \apj, 659, 122

\bibitem[{{Jha} {et~al.}(2006){Jha}, {Kirshner}, {Challis}, {Garnavich},
  {Matheson}, {Soderberg}, {Graves}, {Hicken}, {Alves}, {Arce}, {Balog},
  {Barmby}, {Barton}, {Berlind}, {Bragg}, {Brice{\~n}o}, {Brown}, {Buckley},
  {Caldwell}, {Calkins}, {Carter}, {Concannon}, {Donnelly}, {Eriksen},
  {Fabricant}, {Falco}, {Fiore}, {Garcia}, {G{\'o}mez}, {Grogin}, {Groner},
  {Groot}, {Haisch}, {Hartmann}, {Hergenrother}, {Holman}, {Huchra},
  {Jayawardhana}, {Jerius}, {Kannappan}, {Kim}, {Kleyna}, {Kochanek},
  {Koranyi}, {Krockenberger}, {Lada}, {Luhman}, {Luu}, {Macri}, {Mader},
  {Mahdavi}, {Marengo}, {Marsden}, {McLeod}, {McNamara}, {Megeath}, {Moraru},
  {Mossman}, {Muench}, {Mu{\~n}oz}, {Muzerolle}, {Naranjo}, {Nelson-Patel},
  {Pahre}, {Patten}, {Peters}, {Peters}, {Raymond}, {Rines}, {Schild},
  {Sobczak}, {Spahr}, {Stauffer}, {Stefanik}, {Szentgyorgyi}, {Tollestrup},
  {V{\"a}is{\"a}nen}, {Vikhlinin}, {Wang}, {Willner}, {Wolk}, {Zajac}, {Zhao},
  \& {Stanek}}]{Jha06}
{Jha}, S., {Kirshner}, R.~P., {Challis}, P., {et~al.} 2006, \aj, 131, 527

\bibitem[{{Johansson} {et~al.}(2013){Johansson}, {Thomas}, {Pforr}, {Maraston},
  {Nichol}, {Smith}, {Lampeitl}, {Beifiori}, {Gupta}, \&
  {Schneider}}]{Johansson13}
{Johansson}, J., {Thomas}, D., {Pforr}, J., {et~al.} 2013, \mnras, 435, 1680

\bibitem[{{Kelly} {et~al.}(2015){Kelly}, {Filippenko}, {Burke}, {Hicken},
  {Ganeshalingam}, \& {Zheng}}]{Kelly15}
{Kelly}, P.~L., {Filippenko}, A.~V., {Burke}, D.~L., {et~al.} 2015, Science,
  347, 1459

\bibitem[{{Kelly} {et~al.}(2010){Kelly}, {Hicken}, {Burke}, {Mandel}, \&
  {Kirshner}}]{Kelly10}
{Kelly}, P.~L., {Hicken}, M., {Burke}, D.~L., {Mandel}, K.~S., \& {Kirshner},
  R.~P. 2010, \apj, 715, 743

\bibitem[{{Kessler} {et~al.}(2009{\natexlab{a}}){Kessler}, {Becker}, {Cinabro},
  {Vanderplas}, {Frieman}, {Marriner}, {Davis}, {Dilday}, {Holtzman}, {Jha},
  {Lampeitl}, {Sako}, {Smith}, {Zheng}, {Nichol}, {Bassett}, {Bender}, {Depoy},
  {Doi}, {Elson}, {Filippenko}, {Foley}, {Garnavich}, {Hopp}, {Ihara},
  {Ketzeback}, {Kollatschny}, {Konishi}, {Marshall}, {McMillan}, {Miknaitis},
  {Morokuma}, {M{\"o}rtsell}, {Pan}, {Prieto}, {Richmond}, {Riess}, {Romani},
  {Schneider}, {Sollerman}, {Takanashi}, {Tokita}, {van der Heyden}, {Wheeler},
  {Yasuda}, \& {York}}]{Kessler09}
{Kessler}, R., {Becker}, A.~C., {Cinabro}, D., {et~al.} 2009{\natexlab{a}},
  \apjs, 185, 32

\bibitem[{{Kessler} {et~al.}(2009{\natexlab{b}}){Kessler}, {Bernstein},
  {Cinabro}, {Dilday}, {Frieman}, {Jha}, {Kuhlmann}, {Miknaitis}, {Sako},
  {Taylor}, \& {Vanderplas}}]{Kessler09b}
{Kessler}, R., {Bernstein}, J.~P., {Cinabro}, D., {et~al.} 2009{\natexlab{b}},
  \pasp, 121, 1028

\bibitem[{{Lampeitl} {et~al.}(2010){Lampeitl}, {Smith}, {Nichol}, {Bassett},
  {Cinabro}, {Dilday}, {Foley}, {Frieman}, {Garnavich}, {Goobar}, {Im}, {Jha},
  {Marriner}, {Miquel}, {Nordin}, {{\"O}stman}, {Riess}, {Sako}, {Schneider},
  {Sollerman}, \& {Stritzinger}}]{Lampeitl10}
{Lampeitl}, H., {Smith}, M., {Nichol}, R.~C., {et~al.} 2010, \apj, 722, 566

\bibitem[{{Li} {et~al.}(2011){Li}, {Leaman}, {Chornock}, {Filippenko},
  {Poznanski}, {Ganeshalingam}, {Wang}, {Modjaz}, {Jha}, {Foley}, \&
  {Smith}}]{Li11}
{Li}, W., {Leaman}, J., {Chornock}, R., {et~al.} 2011, \mnras, 412, 1441

\bibitem[{{Maoz} {et~al.}(2014){Maoz}, {Mannucci}, \& {Nelemans}}]{Maoz14}
{Maoz}, D., {Mannucci}, F., \& {Nelemans}, G. 2014, \araa, 52, 107

\bibitem[{{Neill} {et~al.}(2007){Neill}, {Hudson}, \& {Conley}}]{Neill07}
{Neill}, J.~D., {Hudson}, M.~J., \& {Conley}, A. 2007, \apjl, 661, L123

\bibitem[{{Neill} {et~al.}(2009){Neill}, {Sullivan}, {Howell}, {Conley},
  {Seibert}, {Martin}, {Barlow}, {Foster}, {Friedman}, {Morrissey}, {Neff},
  {Schiminovich}, {Wyder}, {Bianchi}, {Donas}, {Heckman}, {Lee}, {Madore},
  {Milliard}, {Rich}, \& {Szalay}}]{Neill09}
{Neill}, J.~D., {Sullivan}, M., {Howell}, D.~A., {et~al.} 2009, \apj, 707, 1449

\bibitem[{{Pike} \& {Hudson}(2005)}]{Pike05}
{Pike}, R.~W., \& {Hudson}, M.~J. 2005, \apj, 635, 11

\bibitem[{{Rest} {et~al.}(2014){Rest}, {Scolnic}, {Foley}, {Huber}, {Chornock},
  {Narayan}, {Tonry}, {Berger}, {Soderberg}, {Stubbs}, {Riess}, {Kirshner},
  {Smartt}, {Schlafly}, {Rodney}, {Botticella}, {Brout}, {Challis}, {Czekala},
  {Drout}, {Hudson}, {Kotak}, {Leibler}, {Lunnan}, {Marion}, {McCrum},
  {Milisavljevic}, {Pastorello}, {Sanders}, {Smith}, {Stafford}, {Thilker},
  {Valenti}, {Wood-Vasey}, {Zheng}, {Burgett}, {Chambers}, {Denneau}, {Draper},
  {Flewelling}, {Hodapp}, {Kaiser}, {Kudritzki}, {Magnier}, {Metcalfe},
  {Price}, {Sweeney}, {Wainscoat}, \& {Waters}}]{Rest14}
{Rest}, A., {Scolnic}, D., {Foley}, R.~J., {et~al.} 2014, \apj, 795, 44

\bibitem[{{Riess} {et~al.}(1996){Riess}, {Press}, \& {Kirshner}}]{Riess96}
{Riess}, A.~G., {Press}, W.~H., \& {Kirshner}, R.~P. 1996, \apj, 473, 88

\bibitem[{{Riess} {et~al.}(1999){Riess}, {Kirshner}, {Schmidt}, {Jha},
  {Challis}, {Garnavich}, {Esin}, {Carpenter}, {Grashius}, {Schild}, {Berlind},
  {Huchra}, {Prosser}, {Falco}, {Benson}, {Brice{\~n}o}, {Brown}, {Caldwell},
  {dell'Antonio}, {Filippenko}, {Goodman}, {Grogin}, {Groner}, {Hughes},
  {Green}, {Jansen}, {Kleyna}, {Luu}, {Macri}, {McLeod}, {McLeod}, {McNamara},
  {McLean}, {Milone}, {Mohr}, {Moraru}, {Peng}, {Peters}, {Prestwich},
  {Stanek}, {Szentgyorgyi}, \& {Zhao}}]{Riess99}
{Riess}, A.~G., {Kirshner}, R.~P., {Schmidt}, B.~P., {et~al.} 1999, \aj, 117,
  707

\bibitem[{{Riess} {et~al.}(2011){Riess}, {Macri}, {Casertano}, {Lampeitl},
  {Ferguson}, {Filippenko}, {Jha}, {Li}, \& {Silverman}}]{Riess11}
{Riess}, A.~G., {Macri}, L., {Casertano}, S., {et~al.} 2011, \apj, 730, 119

\bibitem[{{Rigault} {et~al.}(2013){Rigault}, {Copin}, {Aldering}, {Antilogus},
  {Aragon}, {Bailey}, {Baltay}, {Bongard}, {Buton}, {Canto}, {Cellier-Holzem},
  {Childress}, {Chotard}, {Fakhouri}, {Feindt}, {Fleury}, {Gangler},
  {Greskovic}, {Guy}, {Kim}, {Kowalski}, {Lombardo}, {Nordin}, {Nugent},
  {Pain}, {P{\'e}contal}, {Pereira}, {Perlmutter}, {Rabinowitz}, {Runge},
  {Saunders}, {Scalzo}, {Smadja}, {Tao}, {Thomas}, \& {Weaver}}]{Rigault13}
{Rigault}, M., {Copin}, Y., {Aldering}, G., {et~al.} 2013, \aap, 560, A66

\bibitem[{{Rigault} {et~al.}(2015){Rigault}, {Aldering}, {Kowalski}, {Copin},
  {Antilogus}, {Aragon}, {Bailey}, {Baltay}, {Baugh}, {Bongard}, {Boone},
  {Buton}, {Chen}, {Chotard}, {Fakhouri}, {Feindt}, {Fagrelius}, {Fleury},
  {Fouchez}, {Gangler}, {Hayden}, {Kim}, {Leget}, {Lombardo}, {Nordin}, {Pain},
  {Pecontal}, {Pereira}, {Perlmutter}, {Rabinowitz}, {Runge}, {Rubin},
  {Saunders}, {Smadja}, {Sofiatti}, {Suzuki}, {Tao}, \& {Weaver}}]{Rigault15}
{Rigault}, M., {Aldering}, G., {Kowalski}, M., {et~al.} 2015, \apj, 802, 20

\bibitem[{{Salim} {et~al.}(2007){Salim}, {Rich}, {Charlot}, {Brinchmann},
  {Johnson}, {Schiminovich}, {Seibert}, {Mallery}, {Heckman}, {Forster},
  {Friedman}, {Martin}, {Morrissey}, {Neff}, {Small}, {Wyder}, {Bianchi},
  {Donas}, {Lee}, {Madore}, {Milliard}, {Szalay}, {Welsh}, \& {Yi}}]{Salim07}
{Salim}, S., {Rich}, R.~M., {Charlot}, S., {et~al.} 2007, \apjs, 173, 267

\bibitem[{{Schlafly} \& {Finkbeiner}(2011)}]{Schlafly11}
{Schlafly}, E.~F., \& {Finkbeiner}, D.~P. 2011, \apj, 737, 103

\bibitem[{{Schlegel} {et~al.}(1998){Schlegel}, {Finkbeiner}, \&
  {Davis}}]{Schlegel98}
{Schlegel}, D.~J., {Finkbeiner}, D.~P., \& {Davis}, M. 1998, \apj, 500, 525

\bibitem[{{Scolnic} {et~al.}(2014{\natexlab{a}}){Scolnic}, {Rest}, {Riess},
  {Huber}, {Foley}, {Brout}, {Chornock}, {Narayan}, {Tonry}, {Berger},
  {Soderberg}, {Stubbs}, {Kirshner}, {Rodney}, {Smartt}, {Schlafly},
  {Botticella}, {Challis}, {Czekala}, {Drout}, {Hudson}, {Kotak}, {Leibler},
  {Lunnan}, {Marion}, {McCrum}, {Milisavljevic}, {Pastorello}, {Sanders},
  {Smith}, {Stafford}, {Thilker}, {Valenti}, {Wood-Vasey}, {Zheng}, {Burgett},
  {Chambers}, {Denneau}, {Draper}, {Flewelling}, {Hodapp}, {Kaiser},
  {Kudritzki}, {Magnier}, {Metcalfe}, {Price}, {Sweeney}, {Wainscoat}, \&
  {Waters}}]{Scolnic14b}
{Scolnic}, D., {Rest}, A., {Riess}, A., {et~al.} 2014{\natexlab{a}}, \apj, 795,
  45

\bibitem[{{Scolnic} {et~al.}(2014{\natexlab{b}}){Scolnic}, {Riess}, {Foley},
  {Rest}, {Rodney}, {Brout}, \& {Jones}}]{Scolnic14}
{Scolnic}, D.~M., {Riess}, A.~G., {Foley}, R.~J., {et~al.} 2014{\natexlab{b}},
  \apj, 780, 37

\bibitem[{{Stritzinger} {et~al.}(2011){Stritzinger}, {Phillips}, {Boldt},
  {Burns}, {Campillay}, {Contreras}, {Gonzalez}, {Folatelli}, {Morrell},
  {Krzeminski}, {Roth}, {Salgado}, {DePoy}, {Hamuy}, {Freedman}, {Madore},
  {Marshall}, {Persson}, {Rheault}, {Suntzeff}, {Villanueva}, {Li}, \&
  {Filippenko}}]{Stritzinger11}
{Stritzinger}, M.~D., {Phillips}, M.~M., {Boldt}, L.~N., {et~al.} 2011, \aj,
  142, 156

\bibitem[{{Sullivan} {et~al.}(2006){Sullivan}, {Le Borgne}, {Pritchet},
  {Hodsman}, {Neill}, {Howell}, {Carlberg}, {Astier}, {Aubourg}, {Balam},
  {Basa}, {Conley}, {Fabbro}, {Fouchez}, {Guy}, {Hook}, {Pain},
  {Palanque-Delabrouille}, {Perrett}, {Regnault}, {Rich}, {Taillet}, {Baumont},
  {Bronder}, {Ellis}, {Filiol}, {Lusset}, {Perlmutter}, {Ripoche}, \&
  {Tao}}]{Sullivan06}
{Sullivan}, M., {Le Borgne}, D., {Pritchet}, C.~J., {et~al.} 2006, \apj, 648,
  868

\bibitem[{{Sullivan} {et~al.}(2010){Sullivan}, {Conley}, {Howell}, {Neill},
  {Astier}, {Balland}, {Basa}, {Carlberg}, {Fouchez}, {Guy}, {Hardin}, {Hook},
  {Pain}, {Palanque-Delabrouille}, {Perrett}, {Pritchet}, {Regnault}, {Rich},
  {Ruhlmann-Kleider}, {Baumont}, {Hsiao}, {Kronborg}, {Lidman}, {Perlmutter},
  \& {Walker}}]{Sullivan10}
{Sullivan}, M., {Conley}, A., {Howell}, D.~A., {et~al.} 2010, \mnras, 406, 782

\bibitem[{{Tripp}(1998)}]{Tripp98}
{Tripp}, R. 1998, \aap, 331, 815

\bibitem[{{Tully} \& {Fisher}(1977)}]{Tully77}
{Tully}, R.~B., \& {Fisher}, J.~R. 1977, \aap, 54, 661

\end{thebibliography}






\begin{turnpage}
\begin{deluxetable}{lccccccccccccccccc}
\tablewidth{0pt}
\tablewidth{0pt}
\centering
\setlength{\tabcolsep}{0.06in} 
\tablecolumns{18} 
\tabletypesize{\scriptsize}
\tablecaption{The LSF Step Sample}
\renewcommand{\arraystretch}{1}
\tablehead{Name&Survey\tablenotemark{a}&$z$&
\multicolumn{1}{c}{JLA+PS1} &&
\multicolumn{3}{c}{R11 MLCS2k2 $\Delta M_B^{\textrm{corr}}$}&&
\multicolumn{2}{c}{GALEX data}&&Global&R\tablenotemark{b}&{Dust Corr.}&log($\Sigma_{\textrm{SFR}}$)&P(Ia$\epsilon$)&Cuts\\
\cline{6-8} \cline{10-11}\\
&&& SALT2 $\Delta
M_B^{\textrm{corr}}$&&R$_V=$2.0&R$_V=$2.5&R$_V=3.1$&&Exp.&FUV&&Host Class&&&&\\
\\
&&&&&&&&&(s)&(mag)&&&&&($M_{\odot} kpc^{-2} yr^{-1}$)&(\%)&}
\startdata
010010&PS1&0.100&0.270$\pm$0.113&&\nodata&\nodata&\nodata&&10629&24.90$\pm$0.21&&SF&5.37&N&$-3.072^{+0.081}_{-0.076}$&98&Incl\\*[5pt]
010026&PS1&0.032&0.092$\pm$0.159&&\nodata&\nodata&\nodata&&16222&21.75$\pm$0.04&&SF&1.28&Y&$-2.239^{+0.054}_{-0.050}$&0&\nodata\\*[5pt]
070242&PS1&0.064&0.167$\pm$0.129&&\nodata&\nodata&\nodata&&92341&28.82$\pm$1.09&&SF&35.00&N&$-4.898^{+0.259}_{-0.460}$&100&\nodata\\*[5pt]
10028&SDSS&0.064&-0.102$\pm$0.117&&\nodata&\nodata&\nodata&&3272&25.68$\pm$0.60&&Pa&0.46&N&$-3.712^{+0.194}_{-0.223}$&100&\nodata\\*[5pt]
10805&SDSS&0.044&-0.198$\pm$0.128&&\nodata&\nodata&\nodata&&8006&21.02$\pm$0.04&&SF&0.88&Y&$-1.501^{+0.062}_{-0.053}$&0&\nodata\\*[5pt]
1241&SDSS&0.088&-0.092$\pm$0.108&&\nodata&\nodata&\nodata&&1670&26.04$\pm$1.50&&SF&4.84&N&$-3.321^{+0.254}_{-0.422}$&97&\nodata\\*[5pt]
12779&SDSS&0.079&0.055$\pm$0.122&&\nodata&\nodata&\nodata&&206&24.47$\pm$2.07&&SF&1.91&Y&$-1.983^{+0.384}_{-0.592}$&15&\nodata\\*[5pt]
12781&SDSS&0.083&0.191$\pm$0.119&&\nodata&\nodata&\nodata&&3354&$>$26.43&&Pa&3.34&N&$<-3.670$&100&\nodata\\*[5pt]
12898&SDSS&0.083&0.002$\pm$0.107&&\nodata&\nodata&\nodata&&1627&23.38$\pm$0.24&&SF&0.94&Y&$-2.166^{+0.170}_{-0.295}$&4&\nodata\\*[5pt]
12950&SDSS&0.081&0.078$\pm$0.102&&\nodata&\nodata&\nodata&&4954&22.01$\pm$0.08&&SF&0.55&Y&$-1.817^{+0.112}_{-0.105}$&0&\nodata\\*[5pt]
130308&PS1&0.082&0.037$\pm$0.123&&\nodata&\nodata&\nodata&&4024&24.99$\pm$0.32&&$\sim$SF&0.90&Y&$-2.451^{+0.208}_{-0.329}$&12&Incl\\*[5pt]
17240&SDSS&0.071&-0.159$\pm$0.143&&\nodata&\nodata&\nodata&&3053&$>$27.31&&Pa&4.15&N&$<-3.970$&100&\nodata\\*[5pt]
17258&SDSS&0.088&-0.188$\pm$0.118&&\nodata&\nodata&\nodata&&4130&23.73$\pm$0.20&&SF&0.88&Y&$-1.900^{+0.159}_{-0.251}$&1&\nodata\\*[5pt]
17745&SDSS&0.062&-0.000$\pm$0.117&&\nodata&\nodata&\nodata&&1643&23.34$\pm$0.26&&SF&0.89&Y&$-1.973^{+0.176}_{-0.271}$&2&\nodata\\*[5pt]
18241&SDSS&0.094&0.176$\pm$0.165&&\nodata&\nodata&\nodata&&544&24.19$\pm$0.95&&SF&1.07&Y&$-1.888^{+0.292}_{-0.403}$&5&\nodata\\*[5pt]
19899&SDSS&0.090&-0.048$\pm$0.107&&\nodata&\nodata&\nodata&&2147&25.57$\pm$0.92&&SF&5.88&N&$-3.220^{+0.187}_{-0.334}$&96&\nodata\\*[5pt]
1990af&JRK07&0.050&-0.063$\pm$0.160&&-0.213$\pm$0.170&-0.204$\pm$0.178&-0.205$\pm$0.188&&336&$>$24.09&&Pa&1.77&N&$<-3.230$&100&\nodata\\*[5pt]
1990o&JRK07&0.031&-0.107$\pm$0.150&&-0.071$\pm$0.140&-0.050$\pm$0.144&-0.037$\pm$0.147&&145&22.13$\pm$0.67&&SF&2.95&Y&$-2.156^{+0.300}_{-0.422}$&9&\nodata\\*[5pt]
1990t&JRK07&0.040&-0.048$\pm$0.136&&0.029$\pm$0.194&0.019$\pm$0.203&-0.001$\pm$0.213&&208&23.45$\pm$1.05&&SF&3.92&N&$-3.078^{+0.201}_{-0.383}$&85&\nodata\\*[5pt]
1990y&JRK07&0.039&-0.314$\pm$0.155&&-0.139$\pm$0.259&\nodata&\nodata&&4063&21.41$\pm$0.06&&Pa&1.77&N&$-2.524^{+0.020}_{-0.027}$&0&\nodata\\*[5pt]
1991ag&JRK07&0.014&-0.150$\pm$0.237&&-0.107$\pm$0.199&-0.085$\pm$0.200&-0.068$\pm$0.202&&2301&20.10$\pm$0.05&&SF&1.84&Y&$-2.134^{+0.086}_{-0.072}$&0&\nodata\\*[5pt]
1991s&JRK07&0.056&0.025$\pm$0.129&&0.050$\pm$0.156&0.064$\pm$0.164&0.069$\pm$0.174&&108&23.53$\pm$1.31&&SF&3.36\tablenotemark{$\ast$}&Y&$-2.078^{+0.375}_{-0.526}$&11&\nodata\\*[5pt]
1991u&JRK07&0.033&-0.342$\pm$0.143&&-0.346$\pm$0.174&-0.367$\pm$0.190&-0.398$\pm$0.209&&107&20.67$\pm$0.33&&$\sim$SF&0.35&Y&$-1.593^{+0.204}_{-0.309}$&2&Incl\\*[5pt]
1992ae&JRK07&0.075&-0.182$\pm$0.171&&-0.147$\pm$0.172&-0.086$\pm$0.197&-0.086$\pm$0.215&&224&23.70$\pm$0.89&&Pa&1.84&N&$-2.736^{+0.246}_{-0.316}$&30&\nodata\\*[5pt]
\enddata
\label{table:measurements}
\tablecomments{The full table is available online at \url{http://www.pha.jhu.edu/~djones/lsfstep.html}.}
\tablenotetext{a}{JRK refers to the \citet{Jha07} sample,
  which includes SNe from the CfA1, CfA2, and Calan/Tololo SN surveys \citep{Riess99,Jha06,Hamuy96}.}
\tablenotetext{b}{SN separation from the host galaxy, normalized by
  the SExtractor-measured host galaxy size \citep{Sullivan06}.  We did
  not apply a local dust correction for SNe with R
  $>$ 3, as these are outside the isophotal radius of the
  host.}
\tablenotetext{$\ast$}{Visual inspection found that this SN\,Ia was
  within the isophotal radius of it's host.  A dust correction was applied.}
\end{deluxetable}





\begin{deluxetable}{lcccccccccccc}
\tablewidth{0pt}
\tablewidth{0pt}
\tablecolumns{9} 
\tabletypesize{\scriptsize}
\tablecaption{Local Star Formation Step}
\renewcommand{\arraystretch}{1}
\tablehead{
&
\multicolumn{3}{c}{SALT2} &
\multicolumn{3}{c}{MLCS R$_V$=2.0} &
\multicolumn{3}{c}{MLCS R$_V$=2.5} &
\multicolumn{3}{c}{MLCS R$_V$=3.1}\\
Analysis Change&SNe&$\delta(M^{\textrm{corr}}_{B})_{SF}$&Sig.&
SNe&$\delta(M^{\textrm{corr}}_{B})_{SF}$&Sig.&
SNe&$\delta(M^{\textrm{corr}}_{B})_{SF}$&Sig.&
SNe&$\delta(M^{\textrm{corr}}_{B})_{SF}$&Sig.\\}
\startdata
None&179&0.000$\pm$0.018&0.0$\sigma$ (0.0$\sigma$)&157&0.059$\pm$0.023&2.6$\sigma$ (2.5$\sigma$)&156&0.029$\pm$0.025&1.2$\sigma$ (1.0$\sigma$)&155&0.013$\pm$0.028&0.5$\sigma$ (0.4$\sigma$)\\*[2pt]
$P(A_{FUV})$=1.0$\pm$0.6&179&-0.008$\pm$0.017&-0.4$\sigma$ (-0.4$\sigma$)&157&0.062$\pm$0.023&2.7$\sigma$ (2.6$\sigma$)&156&0.029$\pm$0.025&1.1$\sigma$ (1.0$\sigma$)&155&0.012$\pm$0.028&0.4$\sigma$ (0.4$\sigma$)\\*[2pt]
$P(A_{FUV})$=3.0$\pm$0.6&179&0.005$\pm$0.018&0.3$\sigma$ (0.3$\sigma$)&157&0.060$\pm$0.023&2.6$\sigma$ (2.5$\sigma$)&156&0.031$\pm$0.025&1.3$\sigma$ (1.1$\sigma$)&155&0.016$\pm$0.027&0.6$\sigma$ (0.5$\sigma$)\\*[2pt]
$\Sigma_{SFR}$ boundary = -3.1&179&0.017$\pm$0.018&1.0$\sigma$ (0.9$\sigma$)&157&0.071$\pm$0.023&3.1$\sigma$ (2.9$\sigma$)&156&0.044$\pm$0.025&1.8$\sigma$ (1.6$\sigma$)&155&0.028$\pm$0.027&1.0$\sigma$ (0.9$\sigma$)\\*[2pt]
$\Sigma_{SFR}$ boundary = -2.7&179&-0.005$\pm$0.018&-0.3$\sigma$ (-0.3$\sigma$)&157&0.067$\pm$0.022&3.0$\sigma$ (2.8$\sigma$)&156&0.031$\pm$0.025&1.3$\sigma$ (1.1$\sigma$)&155&0.009$\pm$0.028&0.3$\sigma$ (0.3$\sigma$)\\*[2pt]
1 kpc aper. radius&179&-0.005$\pm$0.018&-0.3$\sigma$ (-0.3$\sigma$)&157&0.051$\pm$0.023&2.2$\sigma$ (2.1$\sigma$)&156&0.018$\pm$0.025&0.7$\sigma$ (0.6$\sigma$)&155&0.005$\pm$0.029&0.2$\sigma$ (0.2$\sigma$)\\*[2pt]
3 kpc aper. radius&179&0.022$\pm$0.018&1.2$\sigma$ (1.2$\sigma$)&157&0.057$\pm$0.024&2.4$\sigma$ (2.3$\sigma$)&156&0.031$\pm$0.025&1.3$\sigma$ (1.1$\sigma$)&155&0.016$\pm$0.027&0.6$\sigma$ (0.5$\sigma$)\\*[2pt]
4 kpc aper. radius&179&0.007$\pm$0.019&0.4$\sigma$ (0.4$\sigma$)&157&0.034$\pm$0.025&1.4$\sigma$ (1.3$\sigma$)&156&0.012$\pm$0.026&0.5$\sigma$ (0.4$\sigma$)&155&-0.001$\pm$0.028&-0.0$\sigma$ (-0.0$\sigma$)\\*[2pt]
Global instead of local SFR&179&-0.001$\pm$0.019&-0.1$\sigma$ (-0.1$\sigma$)&157&-0.013$\pm$0.023&-0.6$\sigma$ (-0.5$\sigma$)&156&0.002$\pm$0.025&0.1$\sigma$ (0.1$\sigma$)&155&0.008$\pm$0.029&0.3$\sigma$ (0.2$\sigma$)\\*[2pt]
2.5$\sigma$-clipping&171&0.005$\pm$0.016&0.3$\sigma$ (0.3$\sigma$)&151&0.047$\pm$0.021&2.2$\sigma$ (2.1$\sigma$)&147&0.046$\pm$0.022&2.1$\sigma$ (1.8$\sigma$)&148&0.039$\pm$0.024&1.6$\sigma$ (1.4$\sigma$)\\*[2pt]
Sys. Error\tablenotemark{a}&&0.004&&&0.007&&&0.014&&&0.014&\\
\enddata
\tablenotetext{a}{The systematic error is computed from the standard
  deviation of each type of variant (e.g. aperture size variants, SFR
  boundary variants, etc.).  The global SFR variant is excluded.}
\label{table:syserrtwo}
\end{deluxetable}

\begin{deluxetable}{lcccccccccccc}
\tablewidth{0pt}
\tablewidth{0pt}
\tablecolumns{9} 
\tabletypesize{\scriptsize}
\tablecaption{Star Formation Dispersion}
\renewcommand{\arraystretch}{1}
\tablehead{
&
\multicolumn{3}{c}{SALT2} &
\multicolumn{3}{c}{MLCS R$_V$=2.0} &
\multicolumn{3}{c}{MLCS R$_V$=2.5} &
\multicolumn{3}{c}{MLCS R$_V$=3.1}\\
Analysis Change & SNe & $\sigma_{\textrm{SF}} - \sigma_{\textrm{passive}}$ & Sig. &
SNe & $\sigma_{\textrm{SF}} - \sigma_{\textrm{passive}}$ & Sig. &
SNe & $\sigma_{\textrm{SF}} - \sigma_{\textrm{passive}}$ & Sig. &
SNe & $\sigma_{\textrm{SF}} - \sigma_{\textrm{passive}}$ & Sig.\\}
\startdata
None&179&-0.013$\pm$0.018&\textbf{-0.7$\sigma$ (-0.7$\sigma$)}&157&-0.033$\pm$0.024&\textbf{-1.4$\sigma$ (-1.2$\sigma$)}&156&-0.053$\pm$0.024&\textbf{-2.2$\sigma$ (-1.8$\sigma$)}&155&-0.086$\pm$0.026&\textbf{-3.3$\sigma$ (-2.5$\sigma$)}\\*[2pt]
$P(A_{FUV})$=1.0$\pm$0.6&179&-0.003$\pm$0.017&\textbf{-0.2$\sigma$ (-0.2$\sigma$)}&157&-0.035$\pm$0.024&\textbf{-1.5$\sigma$ (-1.3$\sigma$)}&156&-0.061$\pm$0.025&\textbf{-2.5$\sigma$ (-2.1$\sigma$)}&155&-0.096$\pm$0.027&\textbf{-3.6$\sigma$ (-2.8$\sigma$)}\\*[2pt]
$P(A_{FUV})$=3.0$\pm$0.6&179&-0.016$\pm$0.018&\textbf{-0.9$\sigma$ (-0.9$\sigma$)}&157&-0.028$\pm$0.024&\textbf{-1.2$\sigma$ (-1.1$\sigma$)}&156&-0.047$\pm$0.024&\textbf{-1.9$\sigma$ (-1.6$\sigma$)}&155&-0.077$\pm$0.026&\textbf{-3.0$\sigma$ (-2.3$\sigma$)}\\*[2pt]
$\Sigma_{SFR}$ boundary = -3.1&179&-0.006$\pm$0.018&\textbf{-0.3$\sigma$ (-0.3$\sigma$)}&157&-0.032$\pm$0.024&\textbf{-1.3$\sigma$ (-1.2$\sigma$)}&156&-0.054$\pm$0.025&\textbf{-2.1$\sigma$ (-1.8$\sigma$)}&155&-0.087$\pm$0.026&\textbf{-3.3$\sigma$ (-2.6$\sigma$)}\\*[2pt]
$\Sigma_{SFR}$ boundary = -2.7&179&-0.017$\pm$0.018&\textbf{-1.0$\sigma$ (-0.9$\sigma$)}&157&-0.031$\pm$0.024&\textbf{-1.3$\sigma$ (-1.1$\sigma$)}&156&-0.053$\pm$0.025&\textbf{-2.2$\sigma$ (-1.8$\sigma$)}&155&-0.087$\pm$0.026&\textbf{-3.3$\sigma$ (-2.6$\sigma$)}\\*[2pt]
1 kpc aper. radius&179&-0.004$\pm$0.018&\textbf{-0.2$\sigma$ (-0.2$\sigma$)}&157&0.024$\pm$0.023&1.0$\sigma$ (0.9$\sigma$)&156&-0.009$\pm$0.025&\textbf{-0.4$\sigma$ (-0.3$\sigma$)}&155&-0.058$\pm$0.027&\textbf{-2.1$\sigma$ (-1.7$\sigma$)}\\*[2pt]
3 kpc aper. radius&179&-0.003$\pm$0.019&\textbf{-0.2$\sigma$ (-0.2$\sigma$)}&157&-0.023$\pm$0.024&\textbf{-0.9$\sigma$ (-0.8$\sigma$)}&156&-0.042$\pm$0.025&\textbf{-1.7$\sigma$ (-1.4$\sigma$)}&155&-0.074$\pm$0.026&\textbf{-2.9$\sigma$ (-2.2$\sigma$)}\\*[2pt]
4 kpc aper. radius&179&-0.016$\pm$0.021&\textbf{-0.8$\sigma$ (-0.8$\sigma$)}&157&-0.016$\pm$0.025&\textbf{-0.6$\sigma$ (-0.6$\sigma$)}&156&-0.029$\pm$0.026&\textbf{-1.1$\sigma$ (-1.0$\sigma$)}&155&-0.057$\pm$0.027&\textbf{-2.1$\sigma$ (-1.7$\sigma$)}\\*[2pt]
Global instead of local SFR&179&-0.038$\pm$0.018&\textbf{-2.1$\sigma$ (-2.1$\sigma$)}&157&0.013$\pm$0.023&0.6$\sigma$ (0.5$\sigma$)&156&0.025$\pm$0.024&1.0$\sigma$ (0.9$\sigma$)&155&0.012$\pm$0.026&0.4$\sigma$ (0.3$\sigma$)\\*[2pt]
2.5$\sigma$-clipping&173&-0.015$\pm$0.017&\textbf{-0.9$\sigma$ (-0.9$\sigma$)}&153&-0.008$\pm$0.022&\textbf{-0.3$\sigma$ (-0.3$\sigma$)}&151&-0.018$\pm$0.023&\textbf{-0.8$\sigma$ (-0.6$\sigma$)}&151&-0.032$\pm$0.025&\textbf{-1.3$\sigma$ (-1.0$\sigma$)}\\*[2pt]
Sys. Error\tablenotemark{a}&&0.003&&&0.013&&&0.016&&&0.022&\\
\enddata
\tablenotetext{a}{The systematic error is computed from the standard
  deviation of each type of variant (e.g. aperture size variants, SFR
  boundary variants, etc.).  The global SFR variant is excluded.}
\label{table:scatter}
\end{deluxetable}

\begin{deluxetable}{lcccccccccccccccc}
\tablewidth{0pt}
\tablewidth{0pt}
\tablecolumns{9} 
\tabletypesize{\scriptsize}
\tablecaption{Star Formation Dispersion with \citet{Kelly15} SFR Boundaries}
\renewcommand{\arraystretch}{1}
\tablehead{
&
\multicolumn{4}{c}{SALT2} &
\multicolumn{4}{c}{MLCS R$_V$=2.0} &
\multicolumn{4}{c}{MLCS R$_V$=2.5} &
\multicolumn{4}{c}{MLCS R$_V$=3.1}\\
$\Sigma_{\textrm{SFR}}$ boundary & SNe & $\sigma_{\textrm{passive}}$&$\sigma_{\textrm{SF}}$ & Sig. &
SNe & $\sigma_{\textrm{passive}}$&$\sigma_{\textrm{SF}}$ & Sig. &
SNe & $\sigma_{\textrm{passive}}$&$\sigma_{\textrm{SF}}$ & Sig. &
SNe & $\sigma_{\textrm{passive}}$&$\sigma_{\textrm{SF}}$ & Sig.\\}
\startdata
-1.7 dex&179&0.127$\pm$0.010&0.118$\pm$0.034&0.3$\sigma$&157&0.145$\pm$0.013&0.141$\pm$0.046&0.1$\sigma$&156&0.193$\pm$0.014&0.138$\pm$0.065&0.9$\sigma$&155&0.198$\pm$0.014&0.265$\pm$0.061&-1.1$\sigma$\\
-1.85 dex&179&0.114$\pm$0.010&0.118$\pm$0.026&-0.1$\sigma$&157&0.146$\pm$0.013&0.129$\pm$0.035&0.5$\sigma$&156&0.169$\pm$0.014&0.171$\pm$0.040&-0.0$\sigma$&155&0.177$\pm$0.014&0.256$\pm$0.044&-1.8$\sigma$\\
\enddata
\label{table:Kelly}
\end{deluxetable}
\clearpage
\end{turnpage}
\end{document}